\documentclass[10pt,journal,compsoc]{IEEEtran}
\ifCLASSOPTIONcompsoc
  \usepackage[nocompress]{cite}
\else
  \usepackage{cite}
\fi
\ifCLASSINFOpdf
   \usepackage[pdftex]{graphicx}
   \graphicspath{{../pdf/}{../jpeg/}}
   \DeclareGraphicsExtensions{.pdf,.jpeg,.png}
\else
   \usepackage[dvips]{graphicx}
   \graphicspath{{../eps/}}
   \DeclareGraphicsExtensions{.eps}
\fi

\graphicspath{{figures/}{pictures/}{images/}{./}} 
\usepackage{microtype}                 
\PassOptionsToPackage{warn}{textcomp}  
\usepackage{textcomp}                  
\usepackage{mathptmx}                  
\usepackage{times}                     
\usepackage{cite}                      
\usepackage{tabu}                      
\usepackage{booktabs}                  
\usepackage{mathtools}
\usepackage{bbm}
\usepackage{enumitem}
\usepackage{verbatim}
\usepackage{amsthm,amsmath}
\usepackage{hyperref}
\usepackage{xcolor}
\usepackage{color}
\usepackage{amssymb}
\usepackage{multirow}
\usepackage{tabularx}
\usepackage{etoolbox}
\usepackage{tikz}
\usepackage{listings}

\newcommand{\revised}[1]{{\color{black} #1}}

\newcommand{\rerevised}[1]{{\color{black} #1}}

\newcommand{\schemaname}{\normalsize\texttt}

\usepackage{hyperref}
\newcommand{\refappendixspec}{\ifx\hideappendix\undefined\ref{appendix:spec}\else A\fi}
\newcommand{\refappendixprompt}{\ifx\hideappendix\undefined\ref{appendix:prompt}\else B\fi}
\newcommand{\refappendixtest}{\ifx\hideappendix\undefined\ref{appendix:test}\else C\fi}

\usepackage{xspace}
\newcommand{\insightRecommendation}{insight recommendation\xspace}
\newcommand{\InsightRecommendation}{Insight recommendation\xspace}
\newcommand{\LEVA}{LEVA\xspace}

\usepackage{url}

\usepackage[framemethod=TikZ]{mdframed}
\definecolor{mycolor}{RGB}{224,215,188}
\definecolor{bgcolor}{RGB}{249,245,233}

\newmdenv[
  backgroundcolor=bgcolor,
  topline=false,
  bottomline=false,
  rightline=false,
  leftline=false,
  skipabove=0.5em,
  skipbelow=0.5em,
  innerleftmargin=0.5em,
  innerrightmargin=0.5em,
  innertopmargin=0.5em,
  innerbottommargin=0.5em,
  singleextra={
    \draw[mycolor, line width=0.5pt] ([yshift=0pt]O) -- ([yshift=0pt]O-|P);
    \draw[mycolor, line width=0.5pt] ([yshift=0pt]P-|O) -- ([yshift=0pt]P);
    \draw[mycolor, line width=0.5pt] ([yshift=1.5pt]O) -- ([yshift=1.5pt]O-|P);
    \draw[mycolor, line width=0.5pt] ([yshift=-1.5pt]P-|O) -- ([yshift=-1.5pt]P);
  },
  secondextra={
    \draw[mycolor, line width=0.5pt] ([yshift=0pt]O) -- ([yshift=0pt]O-|P);
    \draw[mycolor, line width=0.5pt] ([yshift=1.5pt]O) -- ([yshift=1.5pt]O-|P);
  },
  firstextra={
    \draw[mycolor, line width=0.5pt] ([yshift=0pt]P-|O) -- ([yshift=0pt]P);
    \draw[mycolor, line width=0.5pt] ([yshift=-1.5pt]P-|O) -- ([yshift=-1.5pt]P);
  }
]{coloredquotation}

\definecolor{output}{RGB}{222,234,240}
\definecolor{mycoloroutput}{RGB}{177,211,229}
\newmdenv[
  backgroundcolor=output,
  topline=false,
  bottomline=false,
  rightline=false,
  leftline=false,
  skipabove=0.5em,
  skipbelow=0.5em,
  innerleftmargin=0.5em,
  innerrightmargin=0.5em,
  innertopmargin=0.5em,
  innerbottommargin=0.5em,
    singleextra={
    \draw[mycoloroutput, line width=0.5pt] ([yshift=0pt]O) -- ([yshift=0pt]O-|P);
    \draw[mycoloroutput, line width=0.5pt] ([yshift=0pt]P-|O) -- ([yshift=0pt]P);
    \draw[mycoloroutput, line width=0.5pt] ([yshift=1.5pt]O) -- ([yshift=1.5pt]O-|P);
    \draw[mycoloroutput, line width=0.5pt] ([yshift=-1.5pt]P-|O) -- ([yshift=-1.5pt]P);
  },
  secondextra={
    \draw[mycoloroutput, line width=0.5pt] ([yshift=0pt]O) -- ([yshift=0pt]O-|P);
    \draw[mycoloroutput, line width=0.5pt] ([yshift=1.5pt]O) -- ([yshift=1.5pt]O-|P);
  },
  firstextra={
    \draw[mycoloroutput, line width=0.5pt] ([yshift=0pt]P-|O) -- ([yshift=0pt]P);
    \draw[mycoloroutput, line width=0.5pt] ([yshift=-1.5pt]P-|O) -- ([yshift=-1.5pt]P);
  }
]{coloredquotationoutput}

\newcommand{\inlinecodesmall}[1]{{\fontsize{8}{12}\spaceskip=0.5pt\xspaceskip=0.5pt\texttt{\detokenize{#1}}}}

\colorlet{punct}{red!60!black}
\definecolor{delim}{RGB}{20,105,176}
\colorlet{numb}{magenta!60!black}

\lstdefinelanguage{json}{
    basicstyle=\footnotesize\ttfamily,
    numbers=left,
    numberstyle=\color{gray}\footnotesize\ttfamily,
    numbersep=4pt,
    tabsize=2,
    showstringspaces=false,
    breaklines=true,
    xleftmargin=16pt,
    literate=
     *{0}{{{\color{numb}0}}}{1}
      {1}{{{\color{numb}1}}}{1}
      {2}{{{\color{numb}2}}}{1}
      {3}{{{\color{numb}3}}}{1}
      {4}{{{\color{numb}4}}}{1}
      {5}{{{\color{numb}5}}}{1}
      {6}{{{\color{numb}6}}}{1}
      {7}{{{\color{numb}7}}}{1}
      {8}{{{\color{numb}8}}}{1}
      {9}{{{\color{numb}9}}}{1}
      {:}{{{\color{punct}{:}}}}{1}
      {,}{{{\color{punct}{,}}}}{1}
      {\{}{{{\color{delim}{\{}}}}{1}
      {\}}{{{\color{delim}{\}}}}}{1}
      {[}{{{\color{delim}{[}}}}{1}
      {]}{{{\color{delim}{]}}}}{1},
}

\definecolor{myblue}{RGB}{47, 85, 151}

\PassOptionsToPackage{svgnames}{xcolor}
\usepackage{xcolor}
\usepackage{xspace, xpunctuate}
\usepackage{listings}  
\usepackage{enumitem}  
\usepackage{multirow}
\definecolor{keywordcolor}{rgb}{0.56, 0.13, 0.00}
\definecolor{ndkeywordcolor}{rgb}{0.05, 0.46, 0.17}
\definecolor{commentcolor}{rgb}{0.41, 0.64, 0.70}
\definecolor{stringcolor}{rgb}{0.25, 0.44, 0.63}
\lstdefinelanguage{TypeScript}{
  keywords={typeof, new, true, false, catch, function, return, null, catch, switch, var, if, in, while, do, else, case, break, boolean},
  morekeywords={[2]{class, export, throw, implements, import, this}},
  identifierstyle=\color{black},
  sensitive=false,
  comment=[l]{//},
  morecomment=[s]{/*}{*/},
  commentstyle=\color{commentcolor}\ttfamily,
  stringstyle=\color{stringcolor}\ttfamily,
  morestring=[b]',
  morestring=[b]"
}
\lstdefinelanguage{mylang}{
}
\lstdefinestyle{mystyle}{
  basicstyle=\footnotesize\ttfamily,
    numbers=left,
    numberstyle=\color{gray}\footnotesize\ttfamily,
    numbersep=4pt,
  frame=none,
  columns=flexible,
  xleftmargin=10pt,
  aboveskip=-1pt,
  belowskip=0pt,
  language=mylang
}

\def\hidebio{}

\begin{document}

\ifx\hidemain\undefined
    \title{LEVA: Using Large Language Models to Enhance Visual Analytics}
    
    \author{Yuheng~Zhao, Yixing~Zhang, Yu~Zhang, Xinyi~Zhao, Junjie~Wang,\\ Zekai~Shao, Cagatay~Turkay, Siming~Chen
\IEEEcompsocitemizethanks{
\IEEEcompsocthanksitem Yuheng Zhao, Yixing Zhang, Xinyi Zhao, Zekai Shao, Siming Chen are with School of Data Science, Fudan University. E-mail: \{yuhengzhao, xinyizhao19, zkshao19, simingchen\}@fudan.edu.cn, \{yixingzhang23, wangjj23\}@m.fudan.edu.cn. 

\IEEEcompsocthanksitem Yu Zhang is with Department of Computer Science, University of Oxford. E-mail: yuzhang94@outlook.com.

\IEEEcompsocthanksitem Cagatay Turkay is with University of Warwick. 
E-mail: Cagatay.Turkay@warwick.ac.uk.

\IEEEcompsocthanksitem Siming Chen and Yu Zhang are the corresponding authors.
}
\thanks{Manuscript received April 19, 2005; revised August 26, 2015.}}
    
    \markboth{Journal of \LaTeX\ Class Files,~Vol.~14, No.~8, August~2015}%
    {Shell \MakeLowercase{\textit{et al.}}: Bare Demo of IEEEtran.cls for Computer Society Journals}
    
    \IEEEtitleabstractindextext{%
    \begin{abstract}
\revised{Visual analytics supports data analysis tasks within complex domain problems.
However, due to the richness of data types, visual designs, and interaction designs, users need to recall and process a significant amount of information when they visually analyze data.
These challenges emphasize the need for more intelligent visual analytics methods.
}
\revised{Large language models have demonstrated the ability to interpret various forms of textual data, offering the potential to facilitate intelligent support for visual analytics. }
We propose \LEVA, a framework \revised{that uses large language models to enhance users' VA workflows} at multiple stages: onboarding, exploration, and summarization.
To support onboarding, we use large language models to interpret visualization designs and view relationships based on system specifications.
\revised{For exploration, we use large language models to recommend insights based on the analysis of system status and data to facilitate mixed-initiative exploration.}
For summarization, we present a selective reporting strategy to retrace analysis history through a stream visualization and generate insight reports with the help of large language models.
\revised{We demonstrate how \LEVA can be integrated into existing visual analytics systems.
Two usage scenarios and a user study suggest that \LEVA effectively aids users in conducting visual analytics.}
\end{abstract}


    \begin{IEEEkeywords}
    \revised{
    \InsightRecommendation, \revised{mixed-initiative, interface agent}, large language models, visual analytics
    }
    \end{IEEEkeywords}}
    
    \maketitle

    \IEEEdisplaynontitleabstractindextext

    %
    \IEEEpeerreviewmaketitle

    \begin{figure*}[!ht]
  \centering 
  \includegraphics[width=\textwidth]{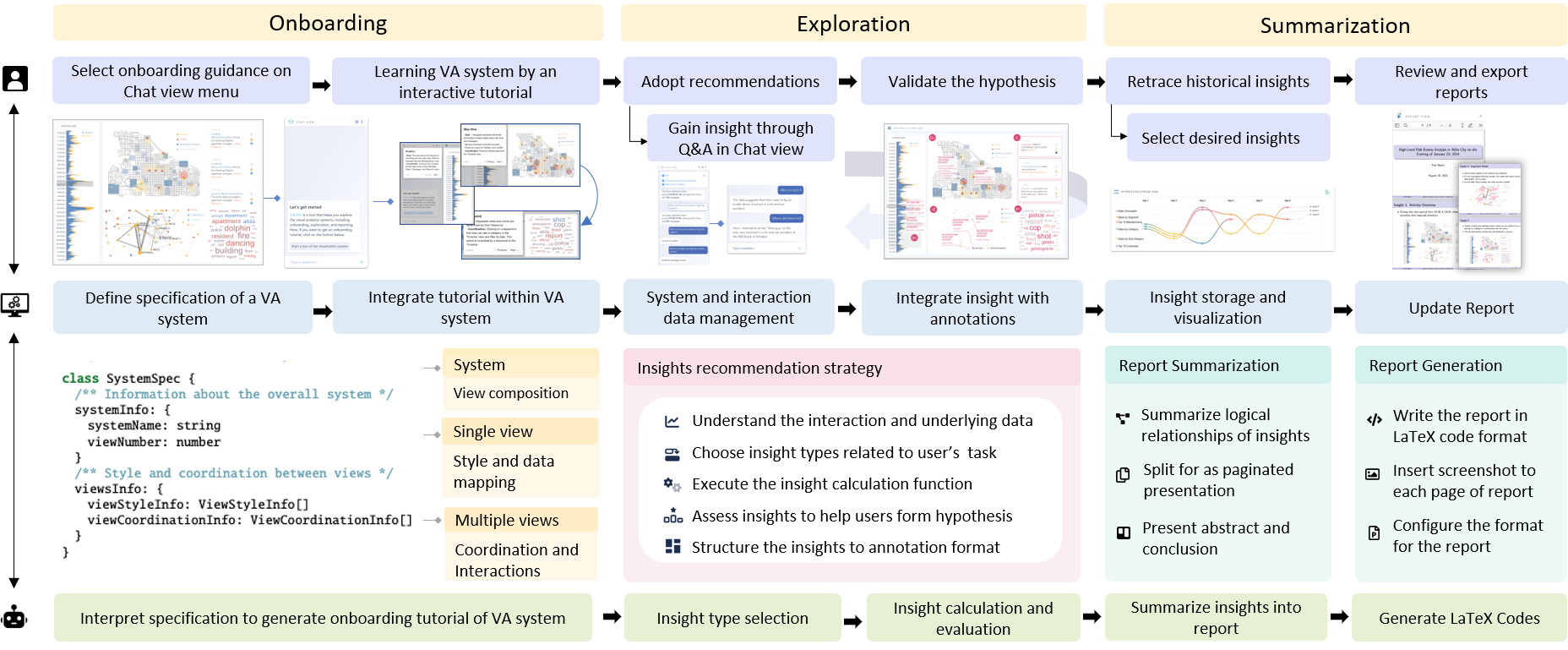}
  \caption{The \LEVA framework proposes strategies for leveraging LLMs to enhance visual analytics workflows, starting from onboarding and exploration to summarization. The architecture (middle) connects analysts (top) and LLMs (bottom) to achieve mixed-initiative exploration through interactive interfaces and guidance strategies of LLMs.}
  \label{fig:framework}
\end{figure*}

\IEEEraisesectionheading{
\section{Introduction}
\label{sec:introduction}}


Visual analytics (VA) combines data analysis techniques with visualizations for effective understanding, reasoning and decision-making on the basis of large and complex datasets~\cite{Keim2008Visual, Kohlhammer2011Solving,Thomas2005Illuminating}.
\revised{However, the VA processes often require significant effort on the side of the user, resulting in a less efficient data interpretation and analysis process~\cite{Wu2023Defence}. Several challenges underpin this inefficiency: a steep learning curve with VA systems' onboarding~\cite{Christina2019Visualization}, a tendency to lose direction in data exploration~\cite{Li2023Diverse}, and the difficulty of summarizing final insights~\cite{Chen2018Supporting}. Such issues emphasize the need for a more intelligent approach to VA.

To specify the challenges further: firstly, the challenge of onboarding stems from unfamiliarity with a VA system, leading to reduced efficiency in grasping visual mappings and interactions~\cite{Ceneda2020Guide, Perez2022Typology}. Secondly, during data exploration, the numerous avenues to analyze data and the lack of structured guidance and direction make it difficult to uncover insights~\cite{Ceneda2017Characterizing}. Thirdly, keeping track of and aggregating key findings is often difficult and summarizing insights is a time-consuming task that demands significant effort and attention to detail~\cite{Liu2020Paths, Callahan2006VisTrails}. These challenges highlight the need for an intelligent framework to support VA processes and foster efficacy across these pivotal stages.} 

To address these problems, researchers focus on enhancing VA through visualization onboarding~\cite{Stoiber2022Perspectives}, interaction recommendation and guidance~\cite{Sperrle2022Lotse, Li2023Diverse,Cao2022Visguide}, and result summarization~\cite{Chen2018Supporting,Sevastjanova2021Visinreport}. However, these methods often do not take advantage of advancements in intelligent algorithms and are not easily adaptable to various VA systems. There is a lack of generalizable intelligent approaches that can work across different VA systems.

The Large Language Models (LLMs) exhibit broader knowledge and problem-solving abilities, making it possible to address the above challenges in the three stages of VA. \revised{Firstly, as LLMs can interpret visualizations' declarative grammar~\cite{Luo2022Natural, Dibia2023LIDA}, we can try to propose the grammar for VA systems and use it as input for LLMs to generate onboarding tutorials. Additionally, LLMs can process various data types~\cite{Openai2023GPT4}, making it possible to recommend insights by analyzing both the system status and underlying data, assisting users' exploration. Finally, LLMs exhibit powerful summarization capabilities~\cite{Goyal2023News}, which may help to summarize the exploration process and generate a report with rich forms.}

In this paper, we propose a framework named \LEVA (\textbf{L}LM-\textbf{E}nhanced \textbf{V}isual \textbf{A}nalytics) that uses LLMs to enhance visual analytics in three stages of the workflow.
\revised{In the onboarding phase, we provide a solution that enables LLMs to interpret visualizations in each of the views and these views' relationships based on a specification of the VA system. This enables the flexible creation of tutorials for various VA systems.
In the exploration phase, we design an \insightRecommendation strategy that guides LLMs in recommending insights based on the understanding of the system, analytical task, user's interaction, and data to facilitate mixed-initiative exploration~\cite{Horvitz1999Principles}.
\rerevised{The methodology incorporates a two-step process, including the selection of insight types and assessment of executed insights.
Additionally, we integrate insights in the original VA system as annotations instead of textual descriptions, making communication between the LLM and end-users more intuitive.}}
In the summarization phase, \LEVA facilitates the user to retrace the analytical history via an interactive stream visualization, allowing the selection of an analytical path for report generation. \rerevised{Our strategy involves a report generation method where the LLMs synthesize visualization images and explanations to produce a report using LaTeX code.}
\revised{
We have developed two integrated systems combining our framework with two original VA systems. The interface includes a chat view, original system view, interaction stream view, and report view.
}
To the best of our knowledge, \LEVA is the first attempt to embed LLMs in complex visual analytics workflows to support users in various analysis stages.
Our main contributions are as follows:

\begin{itemize}[leftmargin=5mm]
    \item We propose a framework, \LEVA, for using LLMs to enhance mixed-initiative exploration in three stages of users' VA workflow. 
    \item \revised{We demonstrate how \LEVA can be implemented in existing VA systems to facilitate visual analytics by enabling a connection between users, interface and LLMs.
    \item We report observations and learnings from two usage scenarios and a user study to demonstrate the effectiveness of our framework.}
\end{itemize}

    \section{Related Work}\label{sec:related_work}

\revised{This section reviews the literature that aims to enhance VA in different stages of users' workflow and the literature using LLMs for VA to examine its abilities.}

\revised{
\subsection{Onboarding in Visual Analytics}
Visualization onboarding is the process of supporting users in reading, interpreting, and extracting information from visual representations of data~\cite{Christina2019Visualization}.
Stoiber et al.~\cite{Stoiber2022Perspectives} found that a VA system may lack low-level information about the data, such as understanding principles of the specific data format, data types, or data structure, which limits data selection and manipulation~\cite{Kohlhammer2011Solving}.
Vaishali et al.~\cite{Vaishali2023Process} found that non-expert users often lack visualization literacy to interpret the data and understand the interactions with and between visualizations in a dashboard.
They pointed out that it is necessary to bridge the knowledge gap between the system and the user's background before exploring the data.
Previous research conducted various onboarding strategies for VA systems. The onboarding form mainly includes textual descriptions, video-based, and step-by-step tours with tooltips and overlays~\cite{Yalcin2016Systematic}. Kwon et al. \cite{Kwon12016Comparative} conducted a user study to compare these methods and found that an interactive tour is better than others with a more engaging experience. 
Previous studies have shown that onboarding tools can help users better understand a VA system. However, these tools are limited in their ability to be easily integrated into different systems. To address this, we proposed an intelligent interactive onboarding method to generate tutorials based on the unified grammar of VA systems.
}

\subsection{Insights Recommendation in Visual Analytics}

Insight recommendations within visualization methodologies have garnered attention in recent VA research, serving as effective instruments to aid users in their analytical tasks. \revised{These methods can broadly be stratified into recommendations encompassing annotations or captions, interactions, and direct visualizations.

A corpus of research has been proposed on generating single visualization~\cite{Qian2021Learning, Hu2019VizML, Moritz2019Formalizing} or multiple-view visualizations~\cite{Deng2023DashBot}. Typically, these studies deal with data table queries, outputting static visualizations. In contrast, our approach adapts to a VA system, producing an enhanced VA system complete with interactive insight annotations. A separate thread of research, exemplified by Lai et al.\cite{Lai2020Automatic} and Liu et al.~\cite{Liu2020AutoCaption}, aims at appending annotations to visualizations. However, their emphasis is limited to single-view visualizations without an overarching framework for comprehensive VA systems.}

\revised{Recognizing the complexity of VA systems, some guidance theories have been proposed~\cite{Ceneda2019Review,Sperrle2022Lotse}. Ceneda et al.~\cite{Ceneda2017Characterizing} characterized interaction guidance along the knowledge gap of the user, the input and output of the guidance generation process, and the degree of guidance that is provided to users. A commonly used implementation method is modeling interaction sequences to predict the next interaction object~\cite{Li2023Diverse}. While this strategy offers a diverse set of suggestions, it needs to collect large amounts of interaction data, grapple with issues of explainability, and be applied to specific VA systems. Instead of relying on interaction data, we mine insights directly from data. We let LLMs understand VA systems and dispatch tasks for computation functions. This approach not only ensures explainability but also fosters adaptability to diverse VA systems.}

\revised{
\subsection{Insights Summarization in Visual Analytics}
In data analysis, the task of translating insights into comprehensive reports necessitates substantial effort from the user for documentation and summarization. Prior research such as DataShot~\cite{Wang2020DataShot} introduced methods for automatic report generation, enabling the derivation of insight-driven reports directly from data along with visuals attached. Li et al.\cite{Li2023Notable} refined this concept by encapsulating data insights within automatically generated notebooks. However, these methods do not support summarization in VA systems. Chen et al.\cite{Chen2018Supporting} innovated a storytelling approach, incorporating recording and editing utilities in VA. VisInReport\cite{Sevastjanova2021Visinreport} offered a tool that fosters manual discourse transcript analysis, and curates reports contingent on user interactions. However, these methods do not automate the visual examination of exploration routes; obtaining insights often requires manual review and refinement. Liu et al.~\cite{Liu2020Paths} designed an analytical graph depicting the journey of exploratory data analysis. Drawing inspiration from this, we fuse a stream visualization to endorse real-time retrospectives. Our novel methodology harnesses LLMs to discern pertinent records, facilitating automatic summation and report creation.
}

\revised{
\subsection{Large Language Models for Visualization}
There has been a rising interest in employing language models for various downstream applications.
Categorizing based on the format of inputs and outputs, LLMs can process code, declarative syntax, natural language, and structured data.

Related work in the visualization domain focuses on understanding or generating visualizations. 
A typical application is using natural language to generate data visualizations through an interface~\cite{Shen2022Towards}.
Moreover, codes~\cite{Maddigan2023Chat2VIS} or declarative grammar for visualizations~\cite{Dibia2023LIDA, Narechania2021NL4DV, Luo2022Natural} is often used as input or output as a simplified form of code. This type of work demonstrates code comprehension of language models and basic knowledge of visualizations.
Inspired by this, we propose to use declarative grammar to represent VA systems, which are essentially combinations of multiple visualizations.

In addition to visualization tasks, Language models can also analyze data~\cite{Zha2023Tablegpt} or act as agents to use tools with APIs~\cite{Qin2023Toolllm}.
Thus, we propose that in addition to using LLMs to process data, some complex computation methods can be used as additional tools, enabling support for a wider range of VA analysis scenarios.
The language model is proficient in generating text with format~\cite{Openai2023GPT4}. For example, Liu et al.~\cite{Liu2022Fill} adopt LLMs for generating semantic input texts based on GUI context. We leverage this prowess to generate and integrate tutorials, insights, and reports to enhance user experience in using VA.
Finally, the language model supports dialogue mode, which allows users to ask and answer questions freely, as well as decompose long tasks to support more complex analysis tasks.

To conclude, the development of LLMs has opened up opportunities to enhance VA. 
Our work is the first trial towards a comprehensive integration of LLMs for intelligent VA.
}

    \section{Motivation}
\label{sec:overview}

\revised{ 
Interpreting and analyzing data with VA often requires significant user effort~\cite{Wu2023Defence}. Various challenges lead to such inefficient analysis, such as the steep learning curve associated with onboarding VA systems~\cite{Christina2019Visualization}, getting sidetracked during data exploration~\cite{Li2023Diverse}, and difficulty summarizing the final insights~\cite{Chen2018Supporting}. These challenges underscore the necessity for a more intelligent approach to VA.
Combining a review of previous studies, we summarized the following design considerations and propose the design requirements aimed at enhancing the user experience in VA.}

\revised{\subsection{Design Considerations}
From our analysis of the VA literature, a recurring concern emerges: despite the advances, users still have to expend excessive effort in various stages of analysis. 
In each of these stages, we identify and dissect specific challenges that intensify the effort users must exert:

\begin{enumerate}[leftmargin=6mm]

\item[\textbf{C1}] \textbf{System onboarding:} 
Before users can extract value from VA, they must first grasp the nuances of the system's design. However, two main challenges hinder this. Firstly, although VA systems are often designed with a target user group in mind, the data and visual encoding displayed in visualization might not always align with the users' background. Ambiguities in data information, such as the meaning of data, structure, types, and transformation, may lead to misinterpretations~\cite{Kohlhammer2011Solving}. Secondly, a lack of visualization literacy among users can complicate the interpretation of visual data, making the initial stage of data analysis more laborious than it should be~\cite{Stoiber2022Perspectives}. This emphasizes the need for a more efficient onboarding process that not only caters to users but also provides clear and intuitive information to smoothen the transition and usage.

\item[\textbf{C2}] \textbf{Insights exploration:}
When embarking on the journey of insight discovery, users are often bogged down by interactively analyzing complex data and visualization, which is time-consuming~\cite{Gotz2009Characterizing}. On top of this, the multi-faceted nature of insights means forming hypotheses requires finding parts of many insights that are relevant to the task and then validating them by interacting to find insights in new states, making the exploration stage more challenging~\cite{Brehmer2013Multi}.
Given these complexities, there emerges a need for VA systems to automatically extract and evaluate insights with users' tasks in mind, ensuring more focused and efficient data analysis.


\item[\textbf{C3}] \textbf{Results summarization:}
Summarizing and reporting findings is a final, yet essential step in VA systems. However, it's a process that is often unduly laborious. Users typically navigate through multiple iterative explorations to validate hypotheses~\cite{Chen2018Supporting}. Yet, not every exploration results in valuable knowledge~\cite{Sevastjanova2021Visinreport}. As a result, analysts frequently spend significant time and effort sifting through, distilling, and manually drafting reports from their interactive exploration outcomes. This highlights the pressing need for automated capturing of findings and allows users to trace back. Furthermore, such automation should provide users with comprehensive illustrated reports~\cite{Li2023Notable}, sparing them the chore of crafting them themselves.
\end{enumerate}
}

\subsection{Design Requirements}
\label{sec:requiremnts}

\revised{
Drawing insights from these design considerations, we have developed a set of targeted design requirements. These requirements are pivotal for enhancing VA, aiming to address the challenges identified earlier.
}
\begin{enumerate}[leftmargin=6mm]
    \item[\textbf{R1}] \textbf{Visual encoding:}
        \revised{To address the barriers of system comprehension and misinterpretations (C1), it's vital that we offer an onboarding tutorial to help users understand how data and visualization are mapped.}
        This could include providing data definitions and encodings.
        
    \item[\textbf{R2}] \textbf{Interaction and coordination:}
        \revised{Considering the challenges users face in discerning the usage of visualizations in a VA system (C1), we should provide an interactive step-by-step tutorial of each view.} The content needs to clarify interactions and the relationship between views.

    \item[\textbf{R3}] \textbf{Insight discovery:}
        \revised{In light of the repetitive analytical tasks users undergo and the depth of domain knowledge required (C2), our framework should provide automated data analysis that helps users discover diversified insights. This may include analysis for different data types.}

    \item[\textbf{R4}] \textbf{Hypothesis formulation and validation:}
        \revised{Given that exploration often occurs in an iterative process (C2), our framework should enable the discovery of insights in constantly updated statuses to facilitate hypothesis formulation and validation.}

    \item[\textbf{R5}] \textbf{Summarization of exploration results:}
        \revised{Considering the labor-intensive nature of the summarization process (C3), our framework should prioritize aiding users in filtering and summarizing their exploration outcomes. This entails the management and visualization of historical insights as well as the capability to generate comprehensive reports.}
\end{enumerate}

\subsection{LLM-Enhanced User Workflow in Visual Analytics}

Our objective is to employ intelligent VA to address the challenges faced by analysts when using VA systems.
However, creating distinct models for each exploration stage, workflow, chart type, data type, and domain is labor-intensive and difficult to scale, requiring general frameworks and models.
Large language models have emerged for various analytical tasks, offering the potential for continuous assistance throughout the workflow.
Thus, we propose a framework that leverages LLMs to enhance visual analytics in three stages of the workflow (Fig.~\ref{fig:vaworkflow}). Based on the analysis of considerations and requirements, we summarize the integration into three stages: \textbf{onboarding}, \textbf{exploration}, insight recommendation
in \textbf{exploration}, and \textbf{summarization} for selective reporting.

\begin{figure}[ht]
    \centering
    \includegraphics[width=\linewidth]{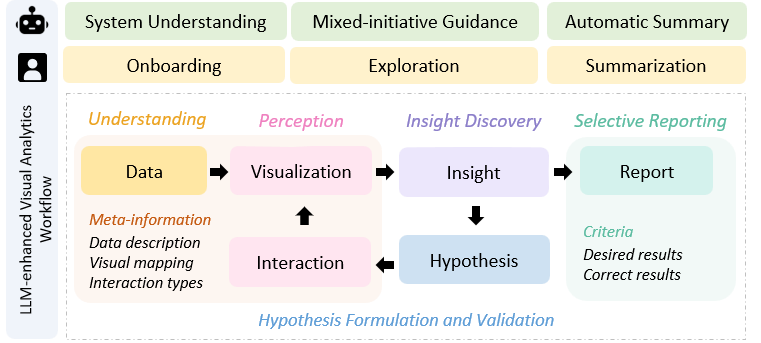}
    \caption{
        The LLM-enhanced visual analytics workflow shows how LLMs contribute to the progress of the analysis.
        The LLMs support visualization understanding, mixed-initiative guidance, and automatic summary while users experience onboarding, exploration, and summarization.
        Onboarding refers to data understanding, visualization, and interaction perception.
        Exploration refers to insight discovery, hypothesis formulation, and validation.
        Summarization refers to selective reporting.
    }
    \label{fig:vaworkflow}
\end{figure}

\revised{To further detail how we intend to tackle these challenges, we outline the roles and purposes of the LLM-enhanced system. The objective of the LLM-enhanced system is to alleviate the challenges users confront in the three stages. The end-users of this enhanced system are analysts or individuals who frequently work with VA systems. On the other hand, the LEVA implementation itself is intended for design and development professionals who aim to create or augment VA systems. The primary users of LEVA are system developers and designers who wish to leverage large language models to enhance the capabilities of their VA tools.}

    \section{LEVA Framework}
\label{sec:framework}


In this section, we will introduce our framework (Fig.~\ref{fig:framework}).
To enhance the VA workflow, we carefully identify the important steps of users in using visual analytics and propose LEVA to boost the analysis efficiency and enrich the insight exploration (Sec.~\ref{sec:overview}.3). The architecture connects analysts and LLMs to achieve mixed-initiative exploration. \revised{In our framework, the original system is a visual analysis system, while our LEVA-enhanced system is added with natural language dialogue, interactive onboarding tutorials, recommended insights, analysis history visualization, and generated reports.
The pipeline of our framework goes through three stages:}
(1) Onboarding: uses LLMs to interpret visualization and views' relationships based on a system specification for users; (2) Exploration: guides LLMs in recommending insights to facilitate user's exploration through the analysis of the system's underlying data. (3) Summarization: allows users to retrace and select their analysis history through visualization and use LLMs to generate insight reports. 


\subsection{VA System Onboarding}
\label{sec:onboarding}

When faced with complex VA systems that contain multiple views, many users experience a steep learning curve. Therefore, we propose an onboarding tutorial generation method, which aims to help users understand two kinds of knowledge in a VA system: visual style and coordination. 

The onboarding can follow two types of approaches: one is bottom-up, i.e., introducing the details first, and the other is top-down, which involves providing an overview initially. Following Tanahashi et al., who demonstrated that top-down is more effective in introducing visualizations~\cite{Tanahashi2016Study}, we take a top-down approach. The tutorial will start from the system-level introduction to the view level. 
To that end, we leverage LLMs' ability to comprehend the system's information and generate tutorials. 
Previous studies demonstrate that declarative grammar of visualization is readily understood by language models\cite{Luo2022Natural, Maddigan2023Chat2VIS}. To encapsulate automatic tutorial generation, we need a unified specification for different VA systems. To this aim, we propose a specification of the VA system designed in line with the top-down approach, which is used as input for LLMs.

\revised{The specification starts from the system-level information \inlinecodesmall{SystemSpec}. We provide the \textit{systemInfo} attribute to detail the system's name and the total number of views it comprises and the \textit{viewsInfo} to describe each view's style and coordination. Examples are available in Appendix~\refappendixspec{}.}

\begin{lstlisting}[language=TypeScript]
class SystemSpec {
  /** Information about the overall system */
  systemInfo: {
    systemName: string
    viewNumber: number
  }
  /** Style and coordination between views */
  viewsInfo: { 
    viewStyleInfo: ViewStyleInfo[]
    viewCoordinationInfo: ViewCoordinationInfo[]
  }
}
\end{lstlisting}

\revised{For the view-level information, we introduced the \inlinecodesmall{viewStyleInfo} to delve into the design of each individual view. This information captures the essence of the VA views through the \inlinecodesmall{viewName} attribute, representing its identity, while the \textit{layers} attribute describes its visual components. Each layer details the mark type (e.g., area), and the encoding attributes that provide insights into how data fields are visually mapped onto axes, color scales, and size scales. Additionally, an optional tooltip specification can be employed to enrich the user's interactive experience by displaying supplementary information upon interaction.}

\begin{lstlisting}[language=TypeScript]
class EncodingInfo {
  field: string;
  type: string;
  description: string;
}
class ViewStyleInfo {
  viewName: string
  /** Information about styles of multiple encodings */
  layers: {
    markType: string
    encoding: {
      x?: EncodingInfo[];
      y?: EncodingInfo[];
      color?: EncodingInfo[];
      size?: EncodingInfo[];
      /** Detail level data field */
      lod?: EncodingInfo[];
    }
    /** Display prompt information for mouse hover.*/
    tooltip?: EncodingInfo[];
  }[]
}
\end{lstlisting}

\revised{Understanding the coordination between different views is pivotal for a coherent exploration experience. 
Thus, we introduced the \inlinecodesmall{ViewCoordinationInfo} into the specification. 
This information bridges the interactions between the source and target views through the \textit{sourceViewName} and \textit{targetViewName} attributes respectively. 
Furthermore, the \textit{coordinationType} attribute delineates the nature of interactions (e.g., filter, brush), and the \textit{interaction} array provides a granular breakdown of user-triggered events and their subsequent ramifications on the target views.}

\begin{lstlisting}[language=TypeScript]
class ViewCoordinationInfo {
  sourceViewName: string
  targetViewName: string | string[]
  /** Coordination type */
  coordinationType: string
  interaction: {
    /** Type of interaction */
    type: string
    /** Interaction's effect on target */
    effect?: {
      /** Action type */
      action: string
      targetViewName: string
      /** Data category for action */
      category: string
      /** Control the result */
      changeby: string
    }
  }[]
}
\end{lstlisting}

\revised{Our specification utilizes a key-value pair approach to determine the choice of fields and their format. This approach could offer flexibility and can be adapted for a variety of visualization types, not just the common visualizations. For example, users may define a customized card using a ``title'' and ``context'' to represent the encoding instead of using ``x'' and ``y''. 
Therefore, for special examples, the definition of key-value pairs is flexible, as long as it makes the LLM understand.
Moreover, the specification not only helps LLMs generate intuitive onboarding tutorials but also aids in understanding the system to assist subsequent stages.}
\subsection{Insight Recommendation in Exploration}
\label{sec:exploration}

Visual exploration can be a challenging task that requires significant effort and expertise, particularly when there are no clear focal points to guide the process. 
Here, we introduce a strategy that channels LLMs towards enhancing insight recommendations. The approach includes three steps: selecting insight types, computing insights, and scoring insights.

\subsubsection{Defining Insight}
\label{sec:insight-definition}

In our framework, insights are considered the basic units in visual exploration. Following previous studies \cite{Ding2019QuickInsights,Wang2020DataShot}, we define insight using four attributes:
\begin{equation}
insight := \langle type, parameters, subject, score \rangle
\end{equation}

\textbf{Insight type:} We first identify the 15 insight \textit{types} based on previous work ~\cite{Ding2019QuickInsights, Stasko2005Information, Bongshin2006Task} including finding extreme, outlier, change point, trend, etc. To cover more complex data types, such as text and graphs, we have added three common types to the list, i.e., text summary and key nodes or key links. More details of the definition of these types are available in the Appendix~\refappendixtest{}. 
Moreover, some insight types may require the analysis of coordinated views within a VA system. 
Therefore, from the perspective of complexity, we define an insight to be either aligned with a single view or across multiple views. For example, the trend of ``sales'' within the ``sales view'', or identifying positive correlations between ``sales'' and ``profit'' in both ``sales view'' and ``profit view''. This distinction will help LLMs understand how to calculate insights in VA systems.


\textbf{Insight parameters:} For each insight type, we use \textit{parameters} to describe the characteristics of an insight, such as the direction of correlation and the location or time of a summarized event. 

\textbf{Insight subject:} The \textit{subject} in our framework defines the data scope to derive an insight, which includes four attributes:
\begin{equation}
subject := \langle subspace, dimension, measure, context \rangle
\end{equation}

Each \textit{subject} corresponds to a subset of data in a view. The values of the \textit{dimension} can be mapped to the x-axis, and the \textit{measure} can be mapped to the y-axis. The \textit{subspace} is a filter of a dataset.  
For example, if a line chart has two products' ``sales'', the \textit{dimension} is the time, \textit{measure} can be ``sales'' or ``profit'', and a \textit{subspace} can be a selected time range or a selected product.
For text data analysis, we can use \textit{measure} to refer to the text data column in a dataset.
In more complex graph data, the \textit{dimension} can refer to the types of entities and relationships within the graph. The \textit{measure} can signify quantifiable properties or attributes related to nodes or edges.

In addition to the data set in the original view, some additional data may also serve as an important data source. For example, derived insights can be reserved to hint at subsequent analysis.
Second, intricate tasks such as finding important nodes associated with particular events surpass the bounds of a conventional \textit{subject} due to the demand for event summaries. 
For these certain requirements, we define the additional data source as \textit{context}. 

\textbf{Insight score:} Different insights are not equally attractive to users. \rerevised{Previous research defines the importance score including impact and significance scores. However, to achieve insight recommendations in the VA system, it is also important to assess what insight types could be used to solve the task. Thus, we consider relevance, impact, and significance scores together to calculate the insight score.}

The significance calculation method is adapted from QuickInsight~\cite{Ding2019QuickInsights}.
Specifically, this score is calculated based on hypothesis testing, which takes a value within the range $[0, 1]$.
Take finding the outstanding number one item in a group as an example. 
The hypothesis is that the data obeys the null hypothesis, i.e., long-tail distribution. The p-value is used to indicate whether the data excludes the maximum that will go against the null hypothesis.
Thus, we get the significance score as $1-p$ ($p$ is the p-value). 
\rerevised{Other examples of calculation methods are introduced in this specification \footnote{\url{https://www.microsoft.com/en-us/research/uploads/prod/2016/12/Insight-Types-Specification.pdf}}.
While our analysis utilizes p-values to assess the significance of insights, we recognize the concerns around the use of p-values. First, the p-value is volatile and does not convey the magnitude of an effect. Second, the p-value is not suitable for all insight types. To address these issues, our scoring mechanism for insights is designed to be adaptable. We suggest incorporating alternative measures such as effect sizes, Cohen's d, Odds Ratio, or the coefficient of determination~\cite{sullivan2012using} for broader applicability. For text analysis, metrics like BLEU and ROUGE or LLMs themselves could serve as more suitable substitutes~\cite{chiang2023large}.
}

\rerevised{To assess the impact score, there are two kinds of aspects that need to be considered. One is the coverage of the data subspace over the entire dataset, and another is the semantics of the data subspace. Previous work defines the impact as the coverage of the data subspace over the entire dataset~\cite{Wang2020DataShot}. Take the social media event analysis (used in Sec.~\ref{sec:case-vast}) as an example, ``microblog'' and ``call center'' are two types of messages that represent two distinct \textit{subspaces}. If the amount of ``microblog'' messages surpasses that of ``call center'', the insights derived from the ``microblog'' data are considered to have a greater impact due to their more extensive coverage. However, from the semantic aspect, the message from the ``call center'' is more impactful because it is more reliable and timely. We assess such impact using LLMs based on their understanding of the dataset and their broad common knowledge of data analysis in different scenarios.}

As users typically engage with a VA system with an analytical task in mind, we use a relevance score to quantitatively measure the congruence between the emergent insight and the overarching analytical task. \rerevised{We let LLMs evaluate the insights by considering how closely the insights align with the task. For instance, in event analysis, the task is to detect risk events and regions in the city. Therefore, analyzing messages to summarize events would score highly on relevance due to its direct connection with the task.} 

Considering three scores jointly, we compute the insight score as $score = \sum_{k \in \{significance, impact, relevance\}} w_k \cdot score_k$.
Here, $w_k$ represents the weights for the significance score, impact score, and relevance score, respectively. Since our purpose of getting data insight is to uncover its patterns, the significance score should be given more weight than the others. Therefore, we empirically set the weights to be 0.5, 0.2, and 0.3, respectively. The weights are adjustable to fit different preferences.

\subsubsection{Recommendation Strategy}
The insights recommendation strategy should be intuitive, adaptive, and relevant to the user's tasks. We address this with the following two-step strategy:

\textbf{Step 1 (Insight type selection):} 
\rerevised{Considering the multitude of possible insight types, we prioritize selecting those that are most relevant to the task.
When the user makes a selection on a specific view, we use LLMs to find relevant insight types according to the user's selection and task and give a relevance score.} The high-relevance score insight types will be used for further calculation. To automatically execute these insight types, LLMs need to schedule the data that compute insight needs (e.g., data tables, column names). The output includes the selected insight types, relevance scores, and the data information for insight calculation.

\textbf{Step 2 (Insight assessment):} \rerevised{This step aims to conduct a comprehensive assessment of insights.} First, the last step's selected insights types will be executed to generate the insights. We use the LLMs to translate the calculated results into complete sentences. Second, we calculate the combined score of insights from three aspects. As we defined before, the significance score is calculated by statistical methods, and the impact and relevance scores are assessed by LLMs. The final output of this step is the insights ranked by the combined score and the corresponding data points in the views for annotation.

After the two steps, the insights should be annotated to the relevant views in the original VA system. Users can constantly interact with the view to gain new insights from LLMs.

\subsection{Summarization of Exploration Results}
\label{sec:summarization}

Following exploration interactions with LLMs, users may wish to embark on a new exploration round, thereby establishing a human-in-the-loop analysis process for the continuous acquisition of novel knowledge. Given that not every interaction stage will yield the desired insights, we advocate for an interactive strategy to filter exploration results and create comprehensive reports, capitalizing on LLM' text summarization abilities.

\revised{
\textbf{Record preservation:}
    Throughout the analytical process, it is imperative to maintain a detailed record of the user's exploration journey, including interactions and insights. \rerevised{We store these elements together because insights and interactions provide key findings that contribute valuable context for report creation. To elaborate, user interactions act as filters that modify the state of the analysis, which defines the subspace of insights. For instance, consider a scenario where a user selects a specific time period. This action prompts the LLM to summarize relevant events in this time period. When presenting the summary of events, it is crucial to provide the time period. This approach ensures that a comprehensive and coherent insight is captured, enhancing the overall understanding and relevance of the insights generated.} Each recorded data set is defined to span \( m \) analysis rounds. Within each round, there are \( n \) distinct steps, representing the user's adopted insights from the LLM or their self-motivated interactions. We can represent the exploration journey as a matrix \( M \) of dimensions \( m \times n \), where each element \( M_{i,j} \) denotes the \( j^{th} \) step in the \( i^{th} \) round.
For each step \( M_{i,j} \), the following details are preserved:

\begin{itemize}[leftmargin=5mm]
    \item \textit{focused view}: a descriptor of which visualization or data view was in focus.
    \item \textit{insights}: the specific insights generated by LLM. If the user interacts by themselves, we record the interaction object.
    \item \textit{screenshots}: a saved screenshot of the focused view crucial for subsequent report illustration.
\end{itemize}

\textbf{Selective reporting:}
To provide users with a cohesive understanding of their analytical journey, our system presents an interactive mechanism. This mechanism graphically illustrates the matrix \( M \), allowing users to retrace the insights for reporting.
Once a user selects a round of data, the curated sequence from \( M \) serves as input for the LLM. Using this structured input, the LLM crafts a report aggregating both the analytical insights and the visualization, capturing the user's exploration journey.
}

    \section{LEVA Implementation}
\label{sec:implementation}


The implementation requirement of our framework includes two parts: the extensions of the original VA system and the LLM-powered components (Fig.~\ref{fig:developing}). The original VA system requires configuration files and data handlers to process user selections and LLM outputs. The LLM-powered components consist of prompt handlers and presentation modules for showcasing outputs across the original system view and three new views.


\begin{figure}[ht]
    \centering
    \includegraphics[width=\linewidth]{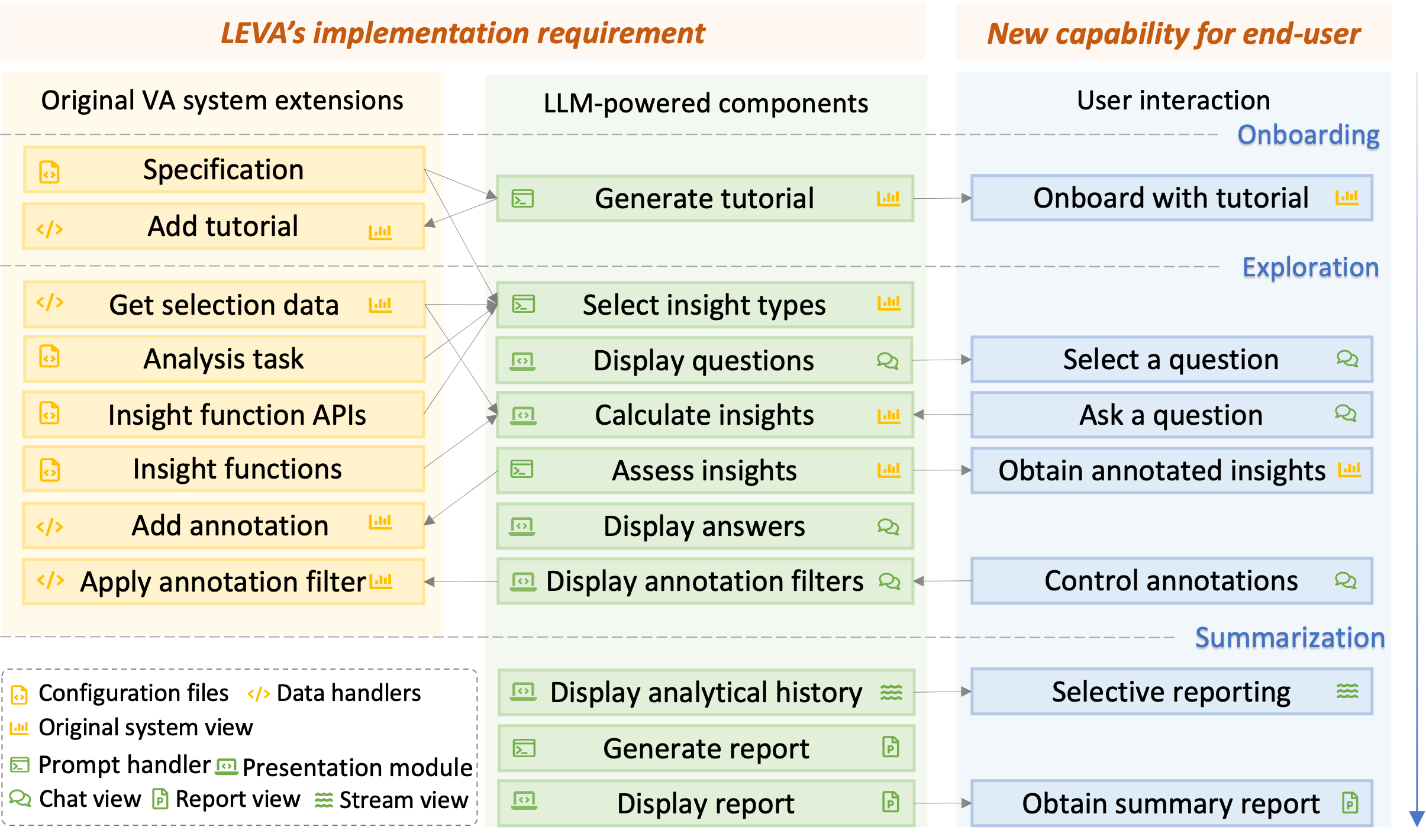}
    \caption{
    The integration overview for augmenting a VA system with
LLM capabilities involves both existing extensions and LLM-powered
components in LEVA’s implementation. The enhancement brings new
capabilities for end-users at three stages.
    }
    \label{fig:developing}
\end{figure}

In the following, we will introduce the efforts to develop the extensions of the original system and the LLM-powered components LEVA provided, including onboarding tutorial generation, insights recommendation, report generation, and the final integrated interface. For prompt templates used, we provide examples in Appendix~\refappendixprompt{}.

\subsection{Extensions of Original System for Integration}
\label{sec:5-components}

\rerevised{To integrate our framework within a VA system, we need to prepare the configuration files and data handlers for the original system.}

\textbf{Specification:} The specification acts as a blueprint for LLMs, allowing them to decode and understand the design and usage of the original VA system.
By comprehending the specifications, LLMs can craft detailed and appropriate tutorials tailored to the VA system. 
\rerevised{Furthermore, as the specification contains the data and coordination information in each view, it can be used to help LLMs explore data by knowing the system state and determine which view and data column to calculate insights from.}

\textbf{Analysis task:} Before recommending insights, it is necessary
to propose a user task to describe in natural language sentences. If
it is not specified, we can also use LLMs to propose the task that
is used to select the relevant insight types. 

\textbf{Insight functions:} We propose a list of functions to calculate insights in different types, e.g., get the outstanding top one and find an outlier. These functions correspond to the insight type defined in Sec. \ref{sec:insight-definition}. The common insight types we give may not be sufficient in some special cases. Thus, developers can add functions according to their domain-specific analysis requirements. Previous taxonomies summarize complex tasks that might be helpful for proposing the computing functions, such as taxonomies for graph analysis~\cite{Bongshin2006Task}, spatial and temporal analysis~\cite{Andrienko2021Theoretical}, social media analysis~\cite{Chen2017Social}.


\rerevised{\textbf{Insight function APIs:} We let LLMs choose the functions from the APIs list to calculate the insight based on their understanding of the system.} To achieve this goal, we need to give the definition of a function. It is better to have two attributes to explain the purpose and outcome of the function: $\langle \textit{name}, \textit{description} \rangle$. For example, to calculate the outstanding number one item in a group of data, the \textit{name} could be ``\textit{get\_outstanding\_top1}'', and the \textit{description} could be ``\textit{Calculate the leading value is significantly higher than all the remaining values}''. 
\rerevised{Providing the APIs of insight functions enables LLMs to choose the suitable analytical method and execute them by completing the parameters.}

\rerevised{
\textbf{Data handlers:}
To support the connection between LLMs and the original system, we need to implement these handlers to capture the changes in the system and add the new features from LLMs.}

\begin{itemize}[leftmargin=5mm]
    \item \textit{Add tutorial:} Receive the tutorials from LLMs and employ a tour guide tool to display the tutorial. 
    \item \textit{Get selection data:} Upon a user interacting with a specific view, the system needs to get the selection data, including the filter and the updated data on each view.
    \item \textit{Add annotation:} The original VA system needs to highlight the insights generated by LLMs on target views with annotations. This function should enable the selection of visual elements, changing their style, and adding annotations around them.
    \item \textit{Apply annotation filter:} Allow the selection of the annotation across the views, such as filtering and deleting.
\end{itemize}

\subsection{LLM-powered Components}

In addition to the above extensions, the enhancement also needs LLM-powered components to support onboarding, exploration, and summarization within the enhanced VA interface.
\subsubsection{Onboarding Tutorial Generation}
\label{sec:5-tutorial}

\revised{
In the onboarding stage, we use LLMs to generate tutorials by inputting the specifications of a VA system. 
To design the output of LLMs, considering the text form of the tutorial will increase the user's reading time, we designed the tutorial as an interactive tour guide. Therefore, the output of LLMs is a list of steps in the tour. Each step within the tour includes two attributes, including \textit{title} and \textit{description}. The value of the \textit{title} is \schemaname{viewName}. The \textit{description} includes visualization \textit{Type}, \textit{Encoding}, and \textit{Coordination}. 
We let LLMs output \textit{description} as HTML format to paraphrase the long text to fewer lines and set font styles for clear observation.
\rerevised{When users select to start the onboarding tour, the tutorial will be triggered and added to the original system to introduce each view.}
The prompt for generating an onboarding tutorial is shown below:
}

\begin{coloredquotation}
    \textbf{Prompt template for onboarding:}\\
    Here are the specifications of a visual analytics system. \\
    \{ \schemaname{specification data} \}\\
    The specification includes the system-level, view-level, and views’ coordination information. You need to introduce each view's style (data meaning, visual mapping) and the relationship between views. Please give your answer in the following format:\\ 
    \{ \schemaname{format requirements} \}
\end{coloredquotation}

\revised{
\subsubsection{Insights Recommendation}
\label{sec:5-recommendation}

We implement the interactive recommendation to undergo two steps of conversation with LLMs. The first step is to select appropriate insight types based on the selection data and analytical task, and the second step is to execute and assess the insight to select the final results.

In the first round, the inputs consist of four components, including specification, the current interaction, analysis task and insight function list.
We added specifications including \inlinecodesmall{view style info} and \inlinecodesmall{views coordination info} to the input. 
The \inlinecodesmall{views coordination info} allows LLMs to know the target views after the user's selection.
The \inlinecodesmall{view style info} contains the data information in the view that will be used as parameters to compose the data for insight calculation.
The current interaction is represented as a triplet: $\langle \textit{viewName}, \textit{dimName}, \textit{value} \rangle$.
\rerevised{If the user selects non-consecutive elements, such as two locations, California and New York, on the map (Sec. \ref{sec:case-tableau}), the current selection will be two triples in an array. Moreover, if the selection from the previous analysis step is not canceled, it will remain together to provide context for further analysis.}
Then, LLMs determine the types of insight that can be analyzed based on the \inlinecodesmall{insight function APIs} and \inlinecodesmall{analytical task} and assess with a relevance score.
In order to execute these insight functions, we define the output of LLMs as a quadruple: $\langle \textit{functionName}, \textit{viewName}, \textit{variableName}, \textit{dimName} \rangle$.
The prompt template for insight type selection is shown below:
}

\begin{coloredquotation}
    \textbf{Prompt template for insight type selection:}\\
    When the user makes an action, the system changes. You should analyze data types of connected views based on the coordination information between views.\\
    \{ \schemaname{current selection} \}\\
    \{ \schemaname{view style info} \}\\
    \{ \schemaname{views coordination info} \}\\
    According to the data info in each view and the analytical task, you should select all suitable analytical functions related to the user's task. You also need to give a relevance score to assess how closely related the insight is to the task.\\ 
    \{ \schemaname{analytical task} \}\\
    \{ \schemaname{insight function APIs} \}\\
    Please give your answer in the following format:\\
    \{ \schemaname{format requirements} \}
\end{coloredquotation}

In the second round, the selected insight functions are executed to get insights and significance scores, which are the \inlinecodesmall{insight calculation results} to be the input for further assessment.
Then, the calculated results are organized into natural language sentences to describe insights by LLMs.
After generating insights and obtaining the significance score, the next step is to assess impact scores. 
\rerevised{
We let LLMs assign an impact score based on the nature of each insight, e.g., potential consequences, urgency and timeliness, and influence on decision-making.
The developer can further modify the definition of impact score to fit specific analysis scenarios.}
The insight will correspond to a triplet: $\langle \textit{viewName}, \textit{dimName}, \textit{value} \rangle$  to locate the insight with highlight effect or annotations on the corresponding Original system view.
The prompt template for insight assessment is shown below:

\begin{coloredquotation}
    \textbf{Prompt template for insight assessment:}\\
    The selected insights are implemented, and the result is returned, including the value and significance score. You also need to give an impact score. You can consider combining your data analysis experience to evaluate from the following aspects: potential consequences, urgency and timeliness, and influence on decision-making. \\
    \{ \schemaname{insight calculation results} \}\\
    Please give your answer in the following format:\\
    \{ \schemaname{format requirements} \}
\end{coloredquotation}

\rerevised{After obtaining the generated insight with the structured format, the \textit{Add annotation} function within the original VA system will be executed to link the data objects from the insights to elements within the views. 
For instance, if the LLM returns insight \inlinecodesmall{{ `viewsName': `Sales|By State', `fieldName': `State/Province', `value': [`California', `New York'], `final_score': 0.5}}, the VA system needs to be able to locate the two points on the map, change the style of the target element (e.g., stroke color), and add an annotation at that location. As we described in Sec.~\ref{sec:insight-definition}, each insight type can correspond to a unique view or multiple views. Thus, if an insight is cross-view, annotations will be added on multiple views as well. 
Considering that it is possible that not all insight types are what the user would like to see, we prefer to use single-view insights to analyze step by step.
}


When using LLMs for data analysis directly, it's crucial to consider the strengths and limitations of the language model. We tested the LLMs' data analysis performance on tabular datasets. The results indicated that their accuracy for basic tasks was relatively low, as detailed in Appendix~\refappendixtest{}. Therefore, for tabular data analysis, we use rule-based methods to guarantee accurate results and use LLMs to invoke these functions based on the understanding of underlying data and the user’s task. However, LLMs demonstrate exceptional competence when analyzing textual data, as evidenced by recent studies~\cite{Goyal2023News, Liu2023Evaluating}. Thus, we use LLMs to analyze these textual data tasks directly. 
Furthermore, if the recommended insights are incomplete or inaccurate, we provide an open-question answering function. 
Users can type follow-up questions to understand the insights in detail. 
Users can type follow-up questions to understand the insights in detail. The LLMs will analyze the underlying data in the current state and return insights
and a highlight of source data, improving the explainability of how
the insights are derived.


We adopt two strategies to address the response speed problem of LLMs. One way is to control the output for efficiency. Considering that some computational tasks, like text summaries, can produce lengthy outputs, it's essential to define the output length in the prompt template to ensure it remains concise. Another way is interactive questions and answers. Before executing insight functions, we adopt an interactive approach where LLMs recommend questions based on the assessment of the relevant insight type derived from the insight selection stage. Users can select a desired question to obtain insights. This alternative strategy can reduce the waiting time to calculate all insights.


In summary, the limitation of input and output influences both the calculation method and the interaction paradigm. Depending on the specific analysis scenario, the above strategy can be fine-tuned to strike the right balance. We also discuss these challenges and future directions in Sec.~\ref{sec:discussion}.

\subsubsection{Report Generation}
\label{sec:5-report}

\revised{
Based on the strategy introduced in Sec. \ref{sec:framework}, we implement the methods in two steps. First, the insights's annotations are saved, and the data is a 5-tuple: (\textit{insight}, \textit{type}, \textit{value}, \textit{viewName}, \textit{imageName}). The combination of analysis round $m$, $n$ and \textit{viewName} is the \textit{image name}, which will be used in LaTeX code. In general, each step obtains the corresponding screenshot according to \textit{viewName}. 
Considering the first and last step preferably needs to give an overview of the entire report, we can set up to capture all views. 
Then, for a round of \inlinecodesmall{historical analysis data}, we let LLMs generate reports in a textual form.
The prompt template for summarizing reports is shown below:
}

\begin{coloredquotation}
    \textbf{Prompt template for report summarization:}\\
    Here is a historical analysis of the system data. The data contains insights that need to be reported:\\
    \{ \schemaname{historical analysis data} \}\\
    Your task is to write an insight report to present these findings. The amount of insight should be equal to the number of steps for given data. Ensure you include both a cover(report title) and a conclusion.\\
    \{ \schemaname{other requirements} \}
\end{coloredquotation}

\revised{
Second, given the exceptional performance of LLMs in code generation, the textual report can be transformed into a LaTeX-formatted presentation report. We can add requirements in the output format to set the report styles.
The final generated reports are presented to the user through an interactive visualization that supports intuitive reading and markup.
The prompt template for LaTeX code generation is shown below:
}
\begin{coloredquotation}
    \textbf{Prompt template for LaTeX code generation:}\\
    Transform the summarized report into LaTeX slides. For each slide, if an insight exhibits a clear hierarchy, segment it using bullet points. Accompany each insight with a screenshot from the system. The filenames for these screenshots can be found in the historical analysis data. Please integrate the following commands for style configuration.\\
    \{ \schemaname{setting requirements} \}
\end{coloredquotation}


\subsubsection{LEVA Interface}

To support the whole framework, we design an interface to bridge users, LLMs, and the underlying data. The interface comprised of four main components: \textit{Chat view}, \revised{\textit{Original system view}}, \textit{Interaction stream view}, and \textit{Report view}, as shown in Fig.~\ref{fig:system}. 


\begin{figure*}[ht]
    \centering
    \includegraphics[width=\linewidth]{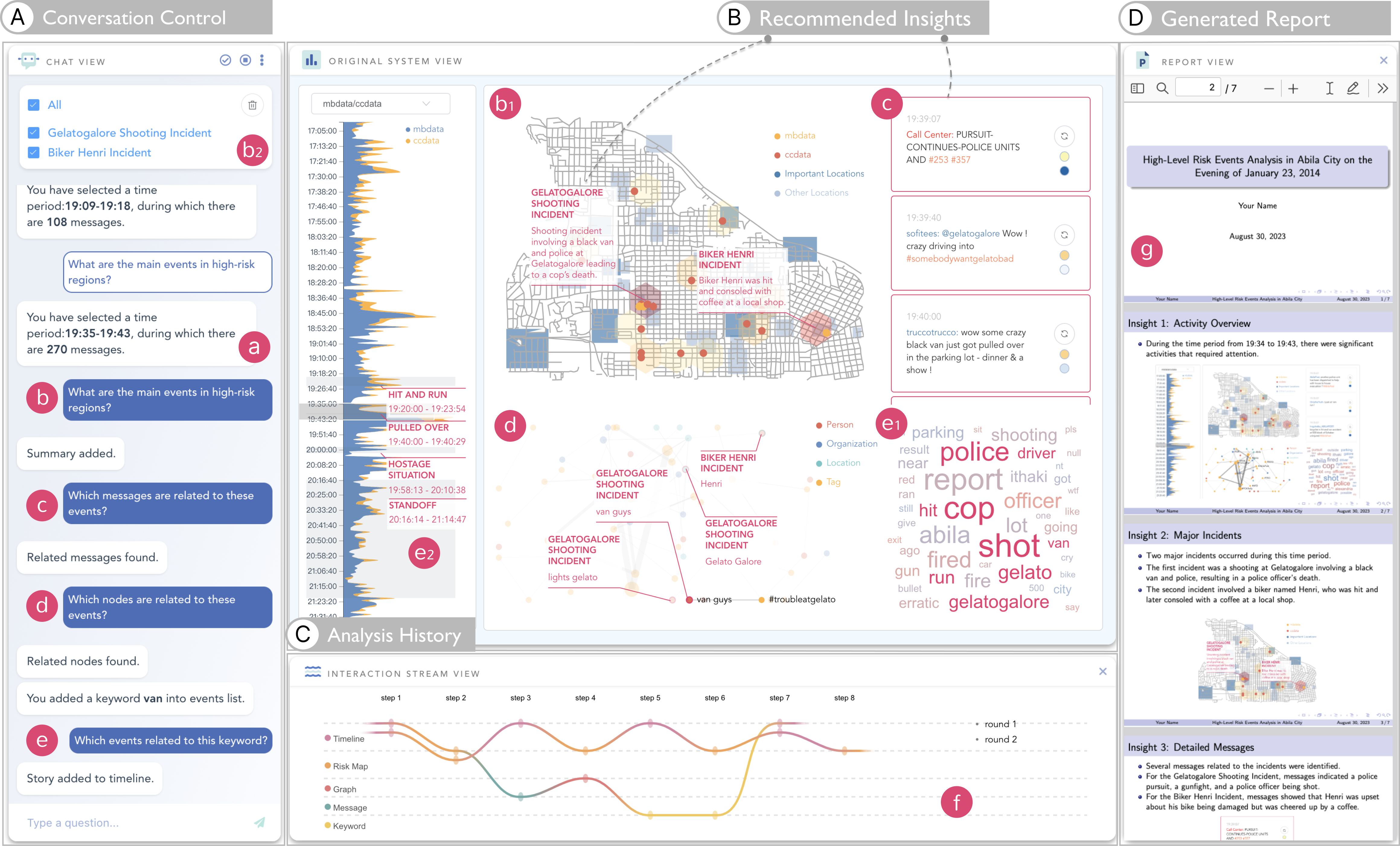}
    \caption{
        An implementation of LEVA comprises of \revised{four components}. Users can communicate with LLMs and control the insight annotations in (a) Chat view; \revised{the recommended insights for next step analysis from LLMs are updated in (b) Original system view;} Users can retrace the interaction history in (d) Interaction stream view; Once a historical analysis path is selected in (d), the generated insight report will display in (e) Report view.}
    \label{fig:system}
\end{figure*}

\revised{
\textbf{Chat view:} To receive feedback and control the entire workflow in our framework, this view serves as an interactive interface where users receive feedback and control other views (Fig. \ref{fig:system}A). During the onboarding stage, the Chat view initiates a tour guide to introduce the original system. \rerevised{When users start on exploration, this view will showcase questions proposed by LLMs. Users can make selections, and the selected questions and insights are recorded and visualized in the Interaction stream view (Fig. \ref{fig:system}C) and to support report generation in Report view (Fig. \ref{fig:system}D).} Moreover, the view allows for open-question answering, letting users engage in fluid conversations with LLMs to clarify doubts or derive new insights. \rerevised{Due to possible wrong formats generated by LLMs, we allow a feedback mechanism to display error messages in the Chat view and make users aware of failed issues.} 

\textbf{Original system view:} The original VA system is combined in this view. To augment the system, LEVA introduces annotations as arguments. These annotations, serving as guiding markers, help users identify interesting data patterns. 
To efficiently manage these annotations, a dedicated control panel in Chat view has been introduced (Fig. \ref{fig:system}b2). It offers filtering capabilities and options to clear all annotations on the view.
\rerevised{Users can interact with the views, prompting LLMs to propose questions based on their selections. After recommending a cross-view insight, if users select an area with annotations in the source view, the system will show the other part of the insights in the target view.}

\textbf{Interaction stream view \& Report view:} The Interaction Stream View stands as a historical ledger, cataloging analytical insights the user obtained from LLMs (Fig. \ref{fig:system}C). Users can hover over the node to retrace details in each step. Each step's analytical insights are automatically saved. Considering that the user may want to stop the current round and start a new one, we provide an \textit{end button} to enable users to end their current analysis round, signaling the system to start a new round of analysis. Moreover, we set the Interaction stream view as hidden by default. Users can open the interaction view in the \textit{menu} at the top of the Chat view when they need to trace back. Upon selecting an interaction path, the Report view (Fig. \ref{fig:system}D) is triggered, presenting a comprehensive report for that round. 
}
    \section{Usage Scenarios}
\label{sec:case-study}

\revised{To evaluate our framework, we demonstrate the LEVA-enhanced VA system in two usage scenarios: one is analyzing multi-facet event data, and the other is analyzing tabular data. \rerevised{We use the OpenAI GPT-4 model in our work.}}

\subsection{Analyzing Multi-faceted Event Data}
\label{sec:case-vast}

To illustrate how LEVA aids users throughout the VA workflow, we opted to reproduce a VA system: the recipient of the IEEE VAST Challenge 2021 Mini-Challenge 3 Award~\cite{Peng2021Mixed}. Our motivation for this choice stems from several compelling reasons. Firstly, this system exemplifies the intricate, human-in-the-loop decision-making tasks inherent to visual analytics. Secondly, it incorporates a representative blend of data types and corresponding visualizations, encompassing text, graph, spatial, and temporal data, which is a typical VA system. Lastly, its recognition as an award winner lends credibility and affirms its representativeness.

The challenge's task is to detect and evaluate public risks in Abila City during the evening of January 23, 2014. The provided data include microblog records and emergency dispatch records from a call center. Thus, the system centers around a comprehensive timeline that serves as the main interaction point to detect event evolution. A message view presents messages from specific time periods. A keyword view displays messages within a selected period, allowing users to hone in on particular topics. A graph view reflects occurrence relations between various entities, such as persons and locations. Finally, the system provides a map view of Abila City to show the message distribution and the risk levels.

In this dataset, we mainly follow the guidelines outlined in Sec.~\ref{sec:implementation}. However, to provide a multi-faceted analysis of the event, we have made some minor adjustments and additional considerations.

\begin{itemize}[leftmargin=5mm]
    \item \textit{insight functions:} We propose some insight types for event analysis. The single-view insight type includes summarizing the events of a certain period in high-risk areas, summarizing the events of a keyword, finding the nodes, messages, and keywords associated with the events, and retrieving values.
    \rerevised{These functions could be related and combined to address a more complex task. For example, summarizing the events could be recommended as the first and then finding the relevant nodes in the graph.} The events serve as a \textit{context} in addition to the graph data, as we defined in Sec.~\ref{sec:insight-definition}. \rerevised{We also define a cross-view insight to analyze multiple views at once, which is to summarize events with temporal and spatial information. The annotations of this insight type will show on both the timeline view and map view.} 
    \item \textit{propose questions:} Considering that textual data analysis may take a long response time, here we let the LLM propose the questions first, and then the user chooses one question on the Chat view to execute the insight functions, as the consideration we described in Sec.~\ref{sec:5-recommendation}.
\end{itemize}

At the beginning, we click the onboarding button to start a tour guide (Fig. \ref{fig:case-onboarding}a). The tutorial introduces the visualization type, visual encoding, and the coordination between views. This guidance leads us to know the meanings of data analyzed in the timeline view (Fig. \ref{fig:case-onboarding}b). We can also obtain that hexagons on the map denote risk levels of the region (Fig. \ref{fig:case-onboarding}d). The system also tells us that all the other views link to the timeline, which renders based on message type and keywords selected from the keyword view (Fig. \ref{fig:case-onboarding}c). Without this guidance, one needs to take more time to explore the system and might build an inaccurate understanding of the system. 

\begin{figure}[ht]
    \centering
    \includegraphics[width=\linewidth]{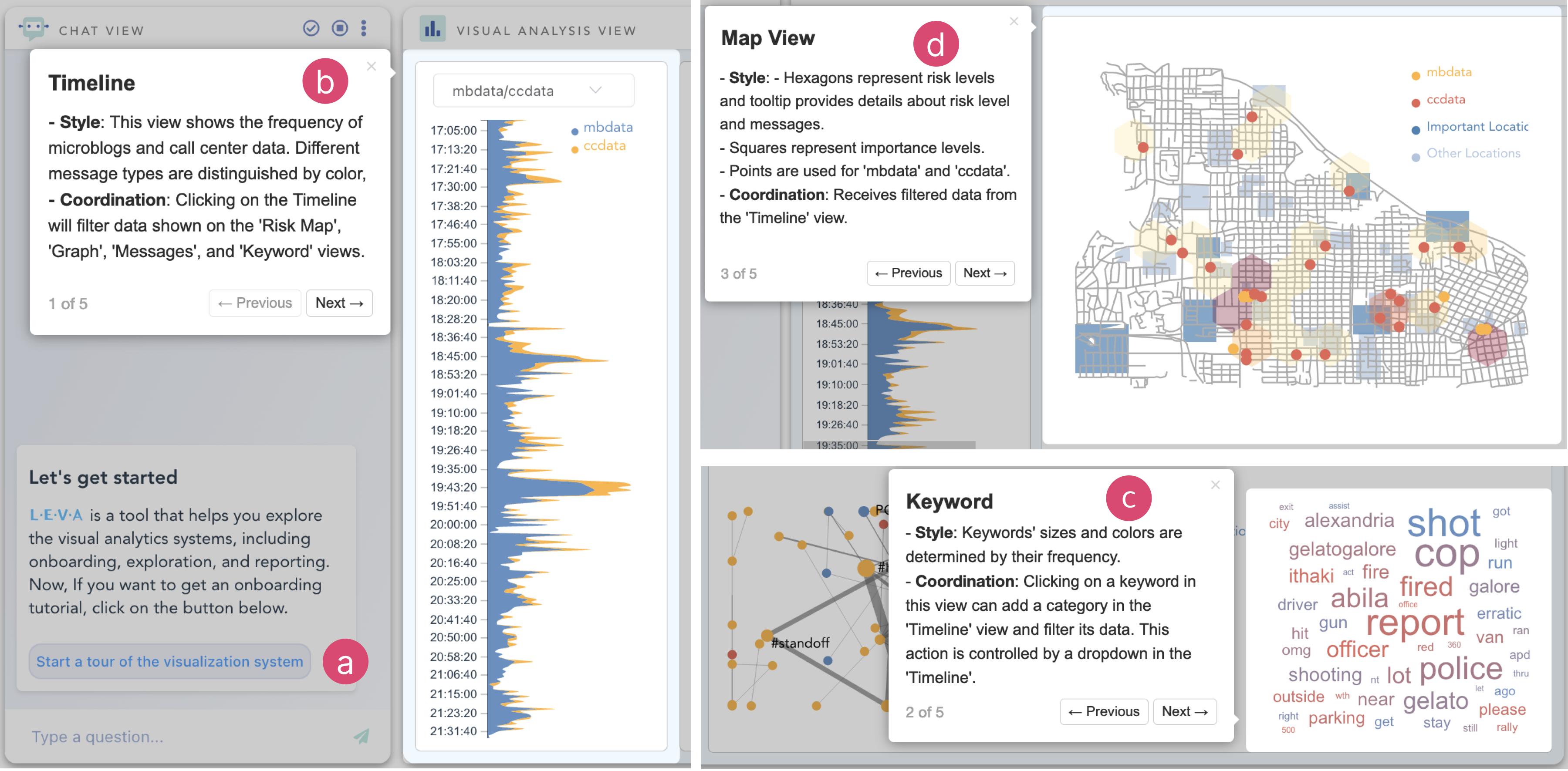}
    \caption{An onboarding tour example of the VAST challenge system. (a) Initiation via the onboarding button, (b) Introductions to data meanings of ``mbdata'' and ``ccdata'', (c) The coordination of keyword view and timeline view based on selected keywords, and (d) The visual encoding of hexagon colors representing risk levels in specific regions.}
    \label{fig:case-onboarding}
\end{figure}

\begin{figure*}[ht]
    \centering
    \includegraphics[width=\textwidth]{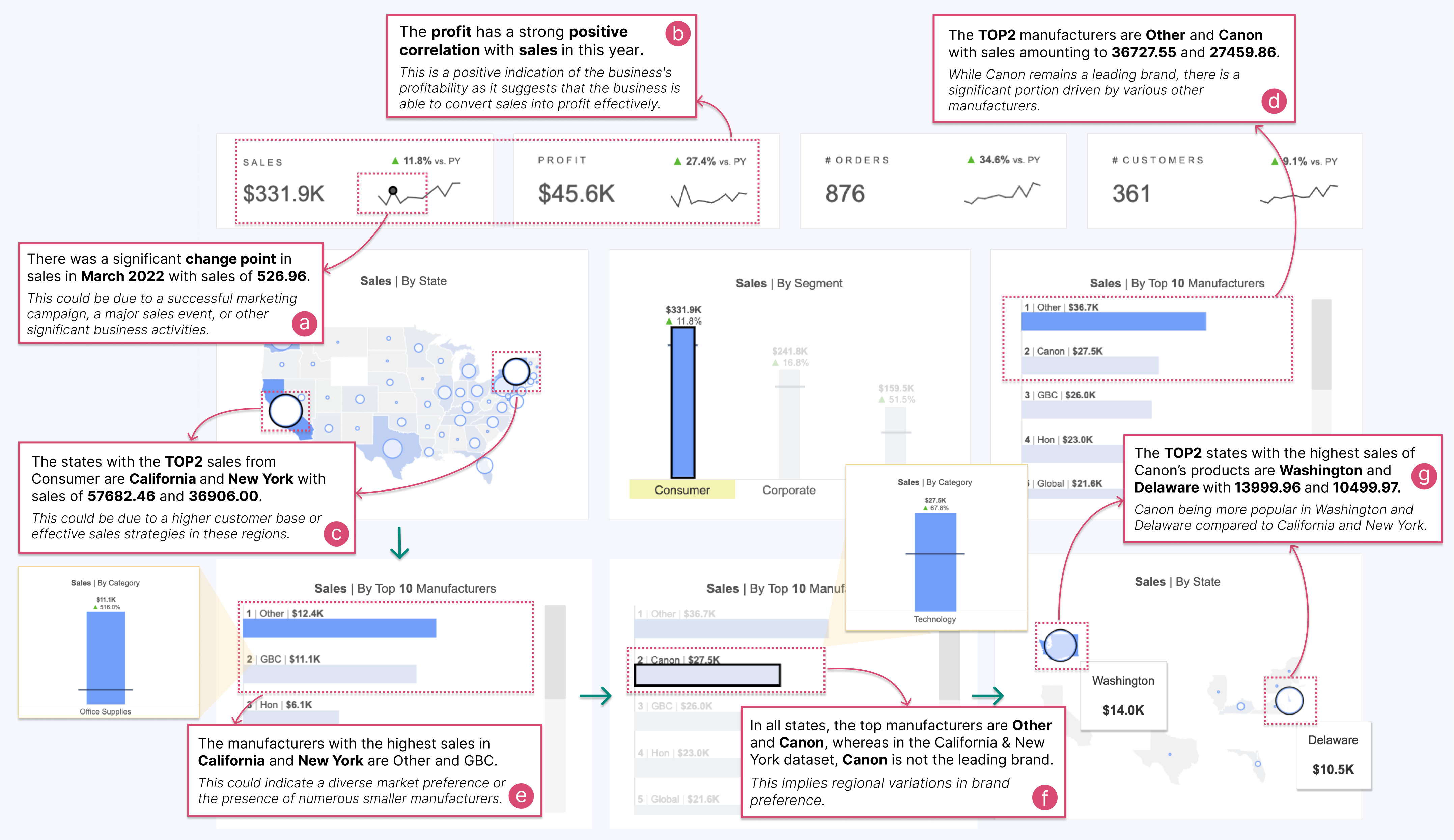}
    \caption{\revised{An example of implementing our framework for tabular data analysis in the exploration stage, which demonstrates the analysis results of each step in a single round (a-d) and the result of multiple rounds (e-g). The annotations represent the insights and potential impact of the LLM output. The analytical process starts with selecting Consumer on Segment view and four exploration steps with LLMs recommendations (a-d). With the help of LLMs, we found that manufacturers with high sales in different regions have a significant distribution of preferences (e-g).}}
    \label{fig:case2}
\end{figure*}

In the next exploration stage, we first look at the timeline as it filters other views. We selected 19:34 to 19:43 since they contain the highest peak. The assistant proposes a question: ``What are the main events in high-risk regions?'' (Fig. \ref{fig:system}b), which is related to our analysis task. Then, the system points out two high-risk regions, summarizes two events (Fig. \ref{fig:system}b1), and figures out the most relevant nodes (Fig. \ref{fig:system}d1) and messages (Fig. \ref{fig:system}c1). By briefly scanning the summaries, we know one high-risk incident is a bicyclist who gets hit but is helped out by people from Brew’ve Been Served. The other is a black van and police shootout in the store. The annotations on the Graph view indicate that the key player in ``Gelatogalore Shooting Incident'' is the ``van guys'', which happens in ``Gelatogalore'', which is a location colored in green. To further know if the black van exits in other time periods, we select "van" from the keyword view (Fig. \ref{fig:system}e1). The assistant catches our action and proposes to analyze the related events with ``van''. We clicked it and four events are summarized along the timeline: ``Hit and run'', ``Pulled over'', ``Hostage situation'', and ``Standoff'' (Fig. \ref{fig:system}e2). After brushing each event on the timeline view, the annotations show up on the map view. Reading the details of each event, we know the whole story of the black van.

In the process of exploration, valuable insights are recorded and can be traced back in the interaction stream graph to avoid forgetting important insights. As incidents related to the black van raise the most public safety concerns, we choose this round to generate the report (Fig. \ref{fig:system}f). The generated report explains each insight and retains the pictures of the exploration process, and summarizes the appropriate title and conclusion page (Fig. \ref{fig:system}g).
\subsection{Analyzing Tabular Data}
\label{sec:case-tableau}

\renewcommand\arraystretch{1.2}
\begin{table*}[ht]
\caption{The questionnaire and the corresponding types, including objective  questions and subjective questions.}
\begin{tabularx}{\textwidth}{lX}
\toprule
\textbf{Type}                                             & \textbf{Specific questions}                                                     \\ 
\midrule
\multirow{3}{*}{\textbf{R1:} Perceptual visual encoding}            & \textbf{O1:}What do the visual encoding and corresponding data mean for the timeline view?                              \\
                                                          & \textbf{S1: }Do you know what each data variaiables means?
   \\
                                                          & \textbf{S2: }Do you understand the meaning of the visual elements?
                                    \\ 
\hline
\multirow{3}{*}{\textbf{R2:} Interaction and Coordination}          & \textbf{O2: }How the timeline view coordinated with other views?                                 \\
                                                          & \textbf{S3: }Are you clear on how to interact in each view?                               \\
                                                          & \textbf{S4: }Are you clear on how the views are related?                                   \\ 
\hline
\multirow{2}{*}{\textbf{R3: }Data Pattern Discovery}                & \textbf{O3: }What high-risk level events occurred in the peak period?                \\
                                                          & \textbf{S5: }Is it easy to get data findings (such as events, key nodes) in these views?  \\ 
\hline
\multirow{3}{*}{\textbf{R4: }Hypothesis Formulation and Validation} & \textbf{O4: }What are the key player and location of the summarized event?             \\
                                                          & \textbf{S6: }Are you clear about the next step analysis for validation?                   \\
                                                          & \textbf{S7: }Do you have easy access to rich hypotheses?                                   \\ 
\hline
\multirow{3}{*}{\textbf{R5: }Summarization of Exploration Results}  & \textbf{O5: }Discover related events of the keyplayer and summarize them into a report.          \\
                                                          & \textbf{S8: }Is it easy to write an analysis report on the interaction results?           \\
                                                          & \textbf{S9: }Are you satisfied with the quality of the report you wrote?                  \\
\bottomrule
\end{tabularx}
\label{subjective questions}
\end{table*}

\revised{To demonstrate the generality of the proposed framework, we apply it to tabular data analysis~\ref{fig:case2}. According to our method design in Sec. \ref{sec:framework}, text data analysis is a task that LLMs are good at, but table data are more suitable for analysis by statistical methods, ensuring accuracy and efficiency. Nonetheless, LLMs remain instrumental, especially in distributing tasks and assessing insights. Thus, this study mainly introduces the exploration stage of tabular data.}

For illustrative purposes, we select a dashboard from Tableau that displays superstore sales data from 2022 \footnote{\url{https://public.tableau.com/app/profile/p.padham/viz/SuperstoreDashboard_16709573699130/SuperstoreDashboard}}. This dashboard features nine distinct views: a choropleth map indicating state-wise sales and five bar charts delineating sales across segments, categories, sub-categories, top 10 manufacturers, and top 10 customers. \revised{Additionally, there are four line charts that trace the trajectories of sales, profits, orders, and customer metrics.}

\revised{Before exploration, we prepare specifications, tasks, insight functions, and data handlers and adjust insights content based on the analytics scenario, following implementation guidelines outlined in Sec.~\ref{sec:5-components}. Some considerations described as below:

\begin{itemize}[leftmargin=5mm]
    \item \textit{specification:} As Tableau dashboards' source files (with file extension .twb) are structured in XML format and contain all the system information that can be extracted and converted to the values in specifications, which allows us to extract specifications automatically. 
    \item \textit{analysis task:} While the dashboards display a variety of data, including ``sales'', ``profit'', ``orders'' and  ``customers'', we initially set a task to focus on analyzing the ``sales'' situation from multiple perspectives.
    \item \textit{insight functions:} We employ fundamental insight functions as referenced in Sec.\ref{sec:insight-definition}, such as outstanding number one, change point, trend, and correlation. Given the breadth of insights possible with tabular data, it is easy for users to overlook connections to prior findings. To address this, we also set to compare current insights with those from the previous step, using the earlier results as \textit{context} for computation, as defined in Sec.\ref{sec:insight-definition}.
    \item \textit{\rerevised{data handlers}:} We integrated the Tableau dashboard into a website using its embedding API~\cite{Tableau2003Embedding}. This API permits ``get selection data'' (Sec.~\ref{sec:5-components}). However, due to API limitations, tutorials and insights are shown in the chat view, and analytical insights are stored accordingly.
    \item \textit{propose explanations:} Considering that in sales analysis, users may need some hint for reason analysis behind the insights. We let the LLM give additional explanations in the final output based on its board knowledge~\cite{Li2023WhyAI}.
\end{itemize}
}

\revised{
In the beginning, we see that the Consumer on sales by segment view was highlighted as the highest value. After selection, the other four bar charts and the map are filtered, and some insights are recommended.
The first recommended insight is ``a significant change point in March 2022'' in the sales line chart (Fig. \ref{fig:case2}a).
The LLMs also suggest potential reasons for the change point could be a successful market campaign or business activities of the superstore.
Then, a recommended insight is a strong positive correlation between profit and sales (Fig. \ref{fig:case2}b).
This finding makes us realize that the strategy launched in February succeeds in turning sales into profits, which can continue to be used in the future.

Beyond time series insights, the recommendations also highlight extreme values, specifically, the top two rankings in states or manufacturers (Fig. \ref{fig:case2}c, d). We notice that the two states with outstanding sales are ``California'' and ``New York''. Consequently, our subsequent analytical focus pivots to these states. Within these jurisdictions, the top two manufacturers identify as ``Other'' and ``GBC'', the latter being a renowned office supply brand (Fig. \ref{fig:case2}e). The below explanation suggests a distinct market inclination towards smaller manufacturers in these regions.

An intriguing insight emerges when comparing the top manufacturers at the state and national levels. While ``Other'' and ``Canon'' dominate sales across most states, ``Canon'' does not maintain this lead in ``California'' and ``New York'' (Fig. \ref{fig:case2}f). This variation accentuates the nuanced manufacturing preferences specific to regions. Motivated by this finding, we move to analyze the distribution of ``Canon'' nationwide. Upon deselecting ``California'' and ``New York'', the recommended insight reveals Washington and Delaware as the leading states and reiterates heightened popularity of ``Canon'' in these regions (Fig. \ref{fig:case2}g). These findings may help potential adjustments to cater to regional sales predilections.
}

    \section{User Study}

\revised{We conducted a user study to evaluate the effectiveness of LEVA for enhancing VA. Specifically, we wanted to verify if LEVA improves users' understanding of the original system, and aids in insights discovery and summarization more efficiently.}

\begin{figure*}[htbp]
\centering
\begin{minipage}[t]{0.33\textwidth}
\centering
\includegraphics[width=6cm]{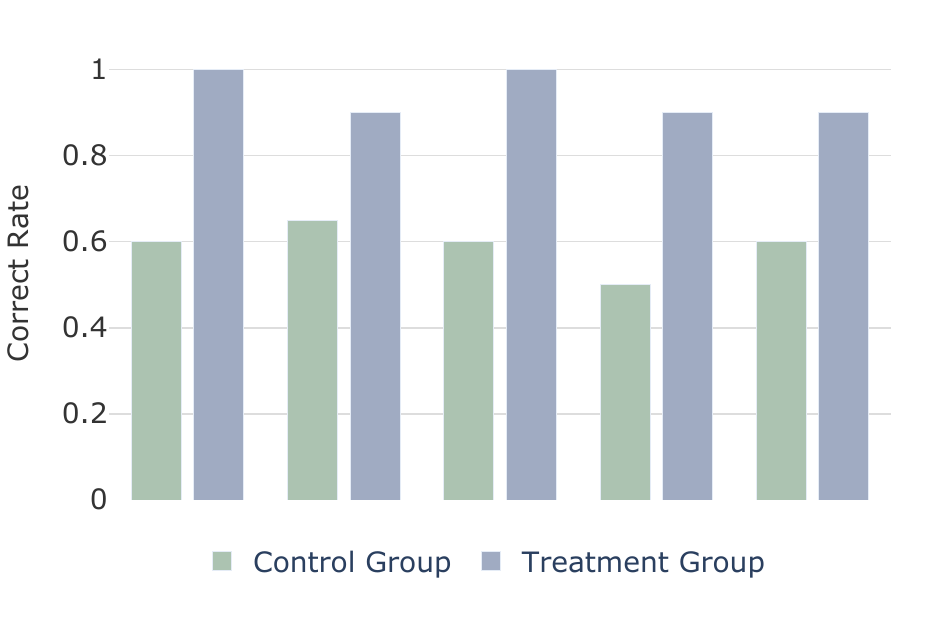}
\end{minipage}
\begin{minipage}[t]{0.33\textwidth}
\centering
\includegraphics[width=6cm]{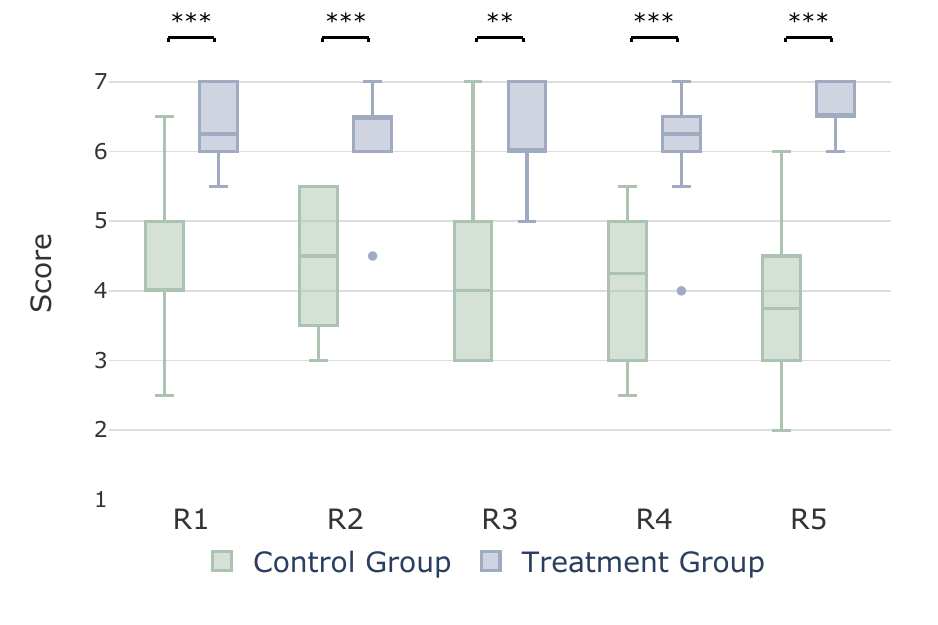}
\end{minipage}
\begin{minipage}[t]{0.33\textwidth}
\centering
\includegraphics[width=6cm]{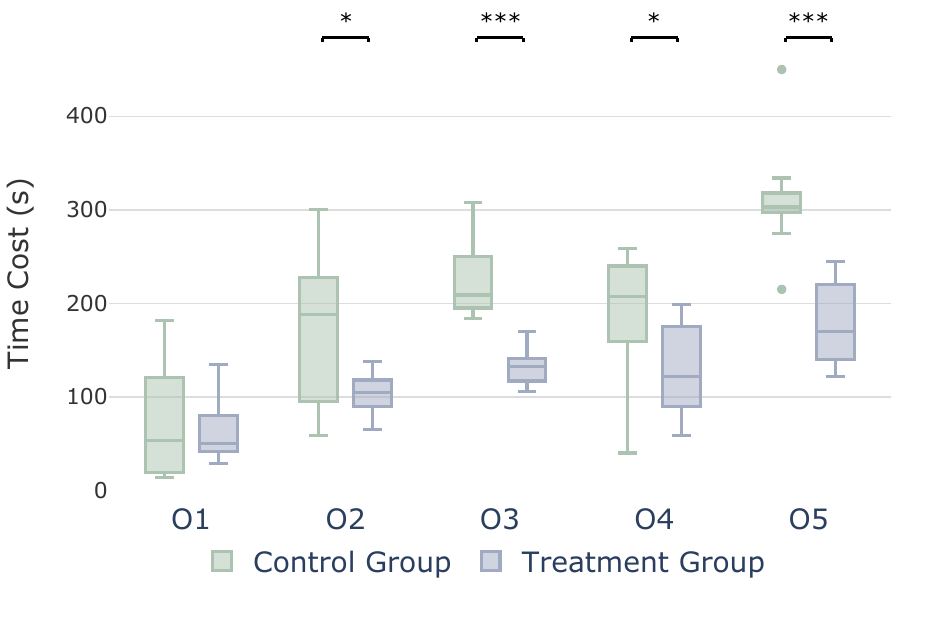}
\end{minipage}
\vspace{-12pt}
\caption{
User Study Results. On the left is the correct rate of five objective questions. In the middle are self-rated scores for subjective questions given by participants. On the right are the answer time for the four timed questions. The number of asterisks (*) in the upper part of the figure indicates the significance level of the test (* : $p<0.05$; ** : $p<0.01$; *** : $p<0.001$). The results suggests our method can improve performance from these three perspectives.
}
\label{result}
\end{figure*}

\subsection{Study Setup}

Here, we introduce the user study by discussing participants and apparatus, questionnaire, procedure, and results analysis.

\textbf{Participants and Apparatus:}
\revised{We recruited 20 participants with backgrounds ranging from computer science, data science, and mathematics to business analysis, ages from 19 to 25 ($\mu = 22.37, \sigma = 1.79$), denoted as P1-P20.
Among them, 4 participants are novices in using VA systems.}
Participants were randomly assigned to two groups, of which 10 participants used the VA system from the VAST challenge system without LEVA’s assistance as the control group and 10 with assistance as the treatment group.
\revised{The studies were all conducted using a monitor with a resolution of 2560 × 1440, along with a mouse and keyboard.}

\textbf{Procedure:} The study was composed of three sessions, beginning with a 10-minute introduction to our framework, the usage of our system, and the original system.
Participants can follow the experimenter to use the system and familiarize themselves with system functions and workflow. 
The formal assessment was conducted using a questionnaire including objective and subjective questions.
Each participant in the two groups was first asked to answer objective questions, followed by subjective questions. 
We also conducted a short interview to collect detailed feedback from the participants.

\textbf{Questionnaire and Measurements:}
\revised{Based on the five requirements (R1-R5) outlined in Section \ref{sec:requiremnts}, we propose five objective questions (O1-O5) and nine subjective questions (S1-S9) to evaluate the effectiveness of our framework in meeting these requirements (Table.~\ref{subjective questions}). Each requirement was tested with at least one objective question and one subjective question. \rerevised{For objective questions, the O1-O4 assessed users' understanding of data and visual mapping, as well as their ability to discover insights from the VA system. 
The O5 required participants to find insights and write a report. The S5-S9 is to verify the effectiveness of LEVA in insight recommendation and summarization.}
Answer times and correct rates were recorded for each question to gauge the influence of LEVA in terms of both efficiency and effectiveness. The correctness assessment for each question is not a strictly binary 0-1 variable but allows for a 0.5 score when answering half of it correctly. For example, for the first question (O1) ``\textit{What do the visual encoding and corresponding data mean for the timeline view?}'', if the user can comprehend the visual encoding but is uncertain about the data meaning, a score of 0.5 is recorded. Finally, the average score of all participants is calculated as the correct rate for this question.} \rerevised{For subjective questions, we included a 7-point Likert scale to allow users to evaluate their level of understanding of the system and whether they encountered any difficulties in gaining data insights and writing analysis reports. 
The control group was asked, ``Do you understand the meaning of the visual elements of each view? Rate your understanding from 1 to 7.'' For the experimental group, the question was modified with the prefix ``With LEVA's assistance'' to gauge the impact of LEVA on understanding. In table~\ref{subjective questions}, we present only the core questions, omitting prefixes and suffixes for brevity.
Among them, S1-S4 are used to verify whether LEVA helps onboarding. Thus, we need to test their understanding of the UI components of the original VA system. The S5-S9 is to verify the effectiveness of LEVA in insight discovery and summarization.
}

\vspace{-10pt}
\subsection{Results and Analysis}
 
\revised{To compare the answer time and subjective scores in two groups, we first conducted the Shapiro-Wilk test in the user study to verify the assumption of normality, ensuring the validity of subsequent t-tests.
The result of the Shapiro-Wilk test confirmed the normal distribution of users' answer times and scores in our samples.} 
After establishing normality, we applied the independent t-test to assess differences in answer times and accuracy between the treatment and control groups, ensuring the reliability of our results. 
We reported our results of the correct rate, subjective scores and time cost.
The detailed result analysis of the user study is presented below.

\textbf{Accuracy:}
We reported the results of objective and subjective measures to assess users' understanding of VA.
The correct rates of five objective questions are shown in Fig.~\ref{result} (left). 
\revised{
For each question, the treatment group had a correct rate of over 85\%, while the control group's correct rate ranged from 50\% to 65\%.
\rerevised{Most participants can distinguish encoding, but as the legend is not specific, they were unclear about the meaning of ``mbdata'' and ``ccdata'' represented by the different colors in the Timeline view} (O1).
Furthermore, in O2, most participants in the control group found it challenging to know the influence of the keyword view on the timeline, while in the treatment group, there were explanations of the interactions between various views. 
In O3, all participants in the treatment group answered correctly, while the control group was only 60\% correct because the two events with higher risk levels were automatically highlighted in the system exploration in the treatment group. 
However, the participants in the control group had to switch between the map and message view and read the text repeatedly.
The O4 yielded a 50\% correctness rate in the control group.
This lower accuracy can be attributed to the fact that a significant portion of participants could only identify the key player involved in the event while struggling to pinpoint the event's location within the intricate graph view. 
In contrast, the treatment group benefited from an automated annotated system that provided clear event location information.
\rerevised{For insight summarization (O5), only a few participants in the control group were able to assemble a more coherent understanding of the event (P12, P15, P16), and they had prior experience in social media VA and cost a considerable amount of time (over 5 minutes). 
In contrast, in the treatment group, only one participant (P4) failed due to their insensitivity to the location name and not noticing the legend.}
}

\textbf{Score:}
\revised{
As shown in Fig.~\ref{result} (middle), the treatment group obtained higher scores than the control group, and the average score is improved by approximately 49.21\%. This indicates that participants perceived LEVA as an improvement over the original VA system in various aspects. 
For R1 and R2, while most participants understood brushing the timeline would filter other views, it was easy to overlook the filtering from the keyword view to the timeline view. Thus, they mostly got half of the score in accuracy but gave a lower subjective score due to the confusion in detailed interactions. This suggests that there was a lack of clear explanations of interactions and coordination in the system, underscoring the significance of effective onboarding. 
The performance in insight discovery varies among participants; only a few people can find some data patterns and form new hypotheses (R3, R4).
\rerevised{For report generation (R5), the treatment group scored significantly higher than the control group in the users' own scores on the ease (S8) and the quality (S9) of generating reports (p \textless 0.001). 
Participants found it difficult to locate previously explored data due to the large amount of information (P18, P19).
The results imply that users typically encounter greater difficulty and produce lower-quality insights when tasked with self-directed exploration and manual report generation compared to the generation of reports through an automated process.}
}



\textbf{Time Cost:}
\revised{
We reported the time costs of the five objective problems in Fig.~\ref{result} (right). 
The results indicate that, with the exception of the first question (O1), the treatment group exhibited shorter answer time compared to the control group ($p<0.05$). For the last four tasks, the average time is reduced by approximately 39.26\% in the treatment group. 
For insight recommendations (O3, O4), the average time is reduced by approximately 41.98\% and 31.27\%.
For the response time of LLMs, summarization events (O3) need more time to return results, about 25s for the long input and output data of O3, but it is still the one with the most time savings.
P2 suggested that \textit{``Even though event summarization took a bit longer than others, it was acceptable given that it saved us even more time than if we had to read and summarize the information ourselves''}.
The LLMs's response time of finding key nodes (O4) is faster than O3, only requiring around 5-6s, improving the average 31.27\% of time.
\rerevised{For the report generation (O5), using the LLM to summarize can save an average of 42.27\% of time compared with no assistance.}
These findings underscore that the utilization of LEVA can substantially diminish the time investment required by users when employing the VA system.
In addition, we observed that LLMs have the probability that the text summary is not comprehensive enough. In the study for P7, the LLM did not clearly summarize who the key player was. However, through the free ask in the chat view, P7 got the answer, which took one minute longer than the average time. This shows that when relying on LLMs for data analysis, it is necessary to provide free questions and answers to ensure the acquisition of detailed information.
}






%

\textbf{Feedback:}
In the final interview, participants were asked about their opinions on the system. 
On one side, participants in the control group offered feedback pertaining to the VA system itself, highlighting issues such as unclear visual elements and legend (P11, P13), confusing color schemes (P20), excessive textual data (P15, P17, P19), and complex interaction between views (P12), no idea where to click and how to explore (P20). On the other side, participants in the treatment group contributed constructive suggestions for improvement:

\rerevised{For onboarding tutorial generation, many participants agreed that onboarding guidance is required, especially for beginners (P9). To improve the tutorial, P4 suggested that ``\textit{The system can further highlight some important keywords in the tutorial}''. 
P7 and P8 recommended incorporating animations, such as arrows, to explain the interactions and relationships between views: ``\textit{Some animations like arrows could be used to explain the interactions and associations between views}.'' This inspires us that the more intuitive tutorial generation could be a future research direction.}

\rerevised{For insight exploration, participants had varied feedback on insight exploration. Some found that the LLM effectively guided their analysis (P4), quickly leading them to valuable insights (P3).
Others mentioned that building on the LLM's analysis, they were inspired to think further and pose new questions by free ask (P7).
However, there were also comments about the improvement of more analytical methods.}
P5 provided a suggestion for improvement: ``\textit{In addition to following the original workflow exploration of the system, like the event analysis, some other tasks, such as starting from a person or spreading relationships, could be considered.}''
\rerevised{The comment points out the need for more types of insight. Therefore, further study could focus on how to bring more domain knowledge and analytical methods to LLMs.} 

\rerevised{For selective report generation, most participants appreciated the reports.  P6 mentioned that ``\textit{not only the comprehensive content with images and texts but also the good formatting}.''}
To improve the report, 
P10 pointed out that ``\textit{If some steps in a stream view could be removed or merged, the generated report would be more useful}.''
\rerevised{The comments demonstrate that they considered our report generation as a convenient and useful function, but further need to improve the log organization and screening.} Additionally, we also collected the comments for scalability. P3 is interested in using LEVA's components as plug-ins for other VA systems. \rerevised{This is a practical suggestion to provide a powerful tool to enhance more VA scenarios we plan to study further. We discuss these valuable suggestions from the user study in the Sec.~\ref{sec:discussion}}



%
%
    \vspace{-5pt}
\section{Discussion}
\label{sec:discussion}
\rerevised{In this section, we discuss the generalizability and the performance of LLMs we observed and highlight the lessons learned from the research and future directions.}

\revised{
\textbf{Generalizability:}
Our framework is generalizable in four aspects. 
First, the system specifications we formulate can be expanded and comprehended by LLMs for tutorial generation. Second, our strategy for recommending insights is adaptable, allowing the LLMs to distribute computational tasks and assess insights in VA systems. Third, our interactive reports generation strategy can be extended to other systems by preserving analysis records. During practical implementations, users can fine-tune these strategies based on the VA tasks and the performance nuances of LLMs. }
Finally, LEVA remains independent of LLMs.
Currently, we integrate LLMs's capabilities to support the exploration of VA workflows.
Although the future appearance of models with other modal inputs saves efforts on engineering implementations of basic information processing, LEVA's strategy for guiding human intelligence model communication remains unchanged.

\revised{\textbf{The performance of LLMs:}}
Despite their immense strengths, LLMs still have several limitations, and there is a lively and ongoing debate on their merits~\cite{Tamkin2021Understanding}.
\rerevised{The first problem is the accuracy.
While one might anticipate that LLMs-enhanced systems will improve as more users interact with them, there are potential risks that LLMs could assist analysts in ways that might not be entirely accurate.
Recent research focuses on using fine-tuned LLMs for tool using~\cite{Qin2023Toolllm}, transforming natural language into code~\cite{Zhang2023Data}, and employing advanced prompting strategies such as self-instruction to compute step-by-step~\cite{Wang2022Self}. 
To enable correct parsing output, we could add an example of constraining the output format of LLMs~\cite{Yousefi2023Incontext}. Moreover, we could also provide an error reporting strategy to make users aware of the unsuccessful response.
We argue that addressing and communicating potential errors remains an exciting and open research challenge and calls for further exploration.}
The second problem is the response time. Our current approach is to control the output length of the LLMs or invoke alternative computation functions that are faster than LLMs. However, controlling the output length may sacrifice the level of detail and depth in the generated output. 
One future direction could be exploring acceleration strategies from the perspective of cache mechanisms~\cite{Doshi2003prefetching}, and predictive analytics might offer speed improvements without compromising the quality of the insights generated.

\rerevised{\textbf{Human-LLM collaboration in open-ended exploration:} In the context of open-ended exploration, the interplay between LLM assistance and human judgment presents a nuanced dynamic. Our user study reveals that the timely questions and insights proposed by LLMs could facilitate efficiency and even prompt users to follow up with their own new questions. However, it is also essential to recognize that users could be over-reliant on LLM guidance. They can thus be steered toward particular directions while missing others. This interplay between guided exploration and autonomous discovery is critical to the design of LLM-supported analytical systems. It warrants careful consideration to balance the benefits of guidance with the freedom of exploration -- a challenge also recognized in the visual analytics guidance literature~\cite{Ceneda2017Characterizing}.

\textbf{Domain knowledge integration for LLMs in specific tasks:}
While ensuring accuracy through the LLM-based insight function invocation mechanisms, there is a need to enhance problem-solving flexibility in specific domain scenarios. The user study indicates that some users with domain analysis backgrounds will have more profound analytical ideas in open-question answering. LLMs need better insight into the calculation and understanding of the analytical task. This capability needs more domain knowledge. There might be two methods: one is tailoring insight types and functions that calculate insights for a particular problem, and another is to employ a fine-tuned LLM, specifically optimized to serve the needs of the domain.~\cite{Jeong2024Finetuning}. Both of which are open challenges for further research.

}

\textbf{Insight recommendation vs. Interaction recommendation:} Our current insight recommendation focuses on extracting essential insights from the underlying data, capitalizing on the inherent patterns and relationships present within the dataset. Another strategy is interaction recommendation, which derives insights from many user interaction data, learning and predicting the next interaction object~\cite{Li2023Diverse}. Such an approach recommended insights with a more substantial contextual relevance. Looking forward, there's potential to integrate the historical interaction data. This merger can pave the way for more intelligent and contextually aligned insight recommendations.

\rerevised{\textbf{Interpretation of user interaction for report generation:} In report generation, both user selections and LLM-generated insights are pivotal. User selections offer a crucial context for the insights generated by the LLM, making it essential to preserve these selections for a complete exploration record. Currently, our approach can describe the actions taken by the user but falls short in interpreting the underlying motivations of these actions, impacting our capacity to provide comprehensive context in the exploration narrative. Typically, these selections are driven by insights users from their observations of data and combining personal knowledge.  Future directions could include mining the related data patterns and combining more domain knowledge to generate more coherent exploration reports.}

\rerevised{\textbf{Narrative-driven report generation:} 
Current methodologies in report generation within our framework primarily focus on compiling logs of exploratory logs into a step-wise report. Looking
ahead, future research could pivot towards employing narrative structures and strategies for the automatic summarization of these logs. One direction could be using LLMs for automatic summation. The enhancements would come in two folds. Firstly, refining the narrative by identifying and reorganizing story pieces based on data relationships like temporal and spatial transitions~\cite{Li2023GeoCamera}, and then crafting coherent explanations~\cite{Zhao2021Chartstory}. Secondly, enhancing data presentation and narrative flow by considering the transitions between narrative segments~\cite{Shi2021Autoclips} with multi-modality expression.
These narrative techniques could provide a more intuitive understanding of the explored results.}

\rerevised{\textbf{LLM-based enhancement vs. Rule-based enhancement:} The advantage of using LLMs is that it enhances flexibility and scalability in aiding various VA scenarios, offering capabilities beyond what a few lines of traditional coding can achieve. LLMs excel in natural language understanding and generate insights across a broad knowledge domain. While traditional coding is precise, it often becomes cumbersome and inflexible when faced with diverse scenarios and evolving user needs. 
Specifically, it would require creating an extensive set of detailed rules to break down various query requests, matching computational modules with data, binding views to different types of insights, utilizing numerous manual templates to introduce the system, describing computation results, and guiding the exploration process. 
}

\textbf{Componentization and plug-in:}
During our user study, we found out that users expect using our framework to assist them in exploring more VA systems with different tasks. Therefore, it is better to offer a toolkit and divide the functionality of LEVA and the views included in the current implementation into components. 
\rerevised{This toolkit will include the LLM-powered components and programmatic interfaces. For developers of the original VA system to integrate with this toolkit, the minimum developing cost is to provide the extensions and APIs to receive the LLM's output and modify the prompts to customize the output format of tutorials or annotations.}
By doing so, we could allow LEVA to be integrated as plugins in different VA systems.
Further, we plan to provide more templates to allow users to customize the information (e.g., specification) entered into the LLM and the desired tutorials, reports, and \insightRecommendation{} from the LLM according to their needs.

\textbf{Specification for VA systems:}
In the extraction of the declarative grammar required in LEVA, we refer to previous work describing basic charts~\cite{Lin2023DMiner} and adding descriptions of the data table, user interaction, and coordination based on the goal of understanding data, view and \insightRecommendation{}. 
\revised{An abstract-level description of data and functionality for VA specifications can further benefit various downstream tasks. Such abstractions can greatly facilitate endeavors like the automatic generation of visualization systems and automated storytelling. \rerevised{As we found the interaction between views could be introduced with more intuitive annotations and animations in the tutorial, the future work can research on how to improve the specification and guide LLMs to generate such tutorial.}}




    \section{Conclusion}
In this study, we introduced \LEVA, a framework that integrates LLMs into VA workflows to achieve intelligent VA. LEVA enhances visual analytics through
three pivotal stages: onboarding, exploration, and summarization.
During the onboarding stage, it interprets visualizations and their
relationships, fostering the creation of dynamic tutorials. In the
exploration stage, our insight recommendation strategy harnesses
LLMs recommend analytical insights based on the interpretation
of the system’s status and data, enriching visual analysis via
annotations. In the summarization stage, LEVA allows users
to revisit and select analytical history, streamlining the report
generation process. Our integration of LEVA with a VA system led
to the development of an interactive interface that fosters a dialogue
between users and LLMs. We conducted two usage scenarios and
a user study to demonstrate the effectiveness of our framework.
\section*{Acknowledgments}
The authors want to thank the reviewers for their suggestions. This
work is supported by Natural Science Foundation of China
(NSFC No.62202105) and Shanghai Municipal Science and Technology Major Project (2021SHZDZX0103), General Program (No. 21ZR1403300), Sailing Program (No. 21YF1402900) and ZJLab.
    \section*{Supplemental Material}

\rerevised{Appendix~\refappendixspec{} introduces the two specifications of the VAST challenge system and the Tableau system. Appendix~\refappendixprompt{} describes the prompt examples we used to generate tutorials, insights, and reports. Appendix~\refappendixtest{} describes the evaluation of LLMs' data analysis capability.}

    \newpage
    \bibliographystyle{abbrv}
    \bibliography{assets/bibs/reference}

\begin{thebibliography}{10}

\bibitem{Andrienko2021Theoretical}
N.~Andrienko, G.~Andrienko, S.~Miksch, H.~Schumann, and S.~Wrobel.
\newblock A theoretical model for pattern discovery in visual analytics.
\newblock {\em Visual Informatics}, 5(1):23--42, 2021.

\bibitem{Brehmer2013Multi}
M.~Brehmer and T.~Munzner.
\newblock A multi-level typology of abstract visualization tasks.
\newblock {\em IEEE Transactions on Visualization and Computer Graphics},
  19(12):2376--2385, 2013.

\bibitem{Callahan2006VisTrails}
S.~P. Callahan, J.~Freire, E.~Santos, C.~E. Scheidegger, C.~T. Silva, and H.~T.
  Vo.
\newblock {VisTrails}: Visualization meets data management.
\newblock In {\em Proceedings of the ACM SIGMOD International Conference on
  Management of Data}, pages 745--747, 2006.

\bibitem{Cao2022Visguide}
Y.~Cao, X.~Li, J.~Pan, and W.~Lin.
\newblock {VisGuide}: User-oriented recommendations for data event extraction.
\newblock In {\em Proceedings of SIGCHI Conference on Human Factors in
  Computing Systems}, pages 1--13, 2022.

\bibitem{Ceneda2020Guide}
D.~Ceneda, N.~Andrienko, G.~Andrienko, T.~Gschwandtner, S.~Miksch,
  N.~Piccolotto, T.~Schreck, M.~Streit, J.~Suschnigg, and C.~Tominski.
\newblock Guide me in analysis: A framework for guidance designers.
\newblock {\em Computer Graphics Forum}, 39(6):269--288, 2020.

\bibitem{Ceneda2017Characterizing}
D.~Ceneda, T.~Gschwandtner, T.~May, S.~Miksch, H.-J. Schulz, M.~Streit, and
  C.~Tominski.
\newblock Characterizing guidance in visual analytics.
\newblock {\em IEEE Transactions on Visualization and Computer Graphics},
  23(1):111--120, 2017.

\bibitem{Ceneda2019Review}
D.~Ceneda, T.~Gschwandtner, and S.~Miksch.
\newblock A review of guidance approaches in visual data analysis: A multifocal
  perspective.
\newblock {\em Computer Graphics Forum}, 38(3):861--879, 2019.

\bibitem{Chen2018Supporting}
S.~Chen, J.~Li, G.~Andrienko, N.~Andrienko, Y.~Wang, P.~H. Nguyen, and
  C.~Turkay.
\newblock {Supporting Story Synthesis}: Bridging the gap between visual
  analytics and storytelling.
\newblock {\em IEEE Transactions on Visualization and Computer Graphics},
  26(7):2499--2516, 2018.

\bibitem{Chen2017Social}
S.~Chen, L.~Lin, and X.~Yuan.
\newblock Social media visual analytics.
\newblock {\em Computer Graphics Forum}, 36(3):563--587, 2017.

\bibitem{chiang2023large}
C.~Chenghan and H.~Lee.
\newblock Can large language models be an alternative to human evaluations?
\newblock {\em arxiv.2305.01937}, 2023.

\bibitem{Deng2023DashBot}
D.~Deng, A.~Wu, H.~Qu, and Y.~Wu.
\newblock {DashBot}: Insight-driven dashboard generation based on deep
  reinforcement learning.
\newblock {\em IEEE Transactions on Visualization and Computer Graphics},
  29(1):690--700, 2023.

\bibitem{Vaishali2023Process}
V.~Dhanoa, C.~Walchshofer, A.~Hinterreiter, H.~Stitz, E.~Groeller, and
  M.~Streit.
\newblock A process model for dashboard onboarding.
\newblock {\em Computer Graphics Forum}, 41(3):501--513, 2022.

\bibitem{Dibia2023LIDA}
V.~Dibia.
\newblock {LIDA:} a tool for automatic generation of grammar-agnostic
  visualizations and infographics using large language models.
\newblock In {\em Proceedings of the Annual Meeting of the Association for
  Computational Linguistics: System Demonstrations}, pages 113--126, 2023.

\bibitem{Ding2019QuickInsights}
R.~Ding, S.~Han, Y.~Xu, H.~Zhang, and D.~Zhang.
\newblock {QuickInsights}: Quick and automatic discovery of insights from
  multi-dimensional data.
\newblock In {\em Proceedings of International Conference on Management of
  Data}, pages 317--332, 2019.

\bibitem{Doshi2003prefetching}
P.~R. Doshi, E.~A. Rundensteiner, and M.~O. Ward.
\newblock Prefetching for visual data exploration.
\newblock In {\em Proceedings of the Eighth International Conference on
  Database Systems for Advanced Applications}, pages 195--202, 2003.

\bibitem{Gotz2009Characterizing}
D.~Gotz and M.~X. Zhou.
\newblock Characterizing users' visual analytic activity for insight
  provenance.
\newblock {\em Information Visualization}, 8(1):42--55, 2009.

\bibitem{Goyal2023News}
T.~Goyal, J.~J. Li, and G.~Durrett.
\newblock News summarization and evaluation in the era of gpt-3.
\newblock {\em arXiv.2209.12356}, 2023.

\bibitem{Horvitz1999Principles}
E.~Horvitz.
\newblock Principles of mixed-initiative user interfaces.
\newblock In {\em Proceedings of the SIGCHI Conference on Human Factors in
  Computing Systems}, pages 159--166, 1999.

\bibitem{Hu2019VizML}
K.~Hu, M.~A. Bakker, S.~Li, T.~Kraska, and C.~Hidalgo.
\newblock {VizML}: A machine learning approach to visualization recommendation.
\newblock In {\em Proceedings of the SIGCHI Conferenceon Human Factors in
  Computing Systems}, pages 1--12, 2019.

\bibitem{Jeong2024Finetuning}
C.~Jeong.
\newblock Fine-tuning and utilization methods of domain-specific llms.
\newblock {\em arxiv.2401.02981}, 2024.

\bibitem{Keim2008Visual}
D.~Keim, G.~Andrienko, J.-D. Fekete, C.~G{\"o}rg, J.~Kohlhammer, and
  G.~Melan{\c{c}}on.
\newblock {\em {Visual Analytics}: Definition, process, and challenges}.
\newblock Springer, 2008.

\bibitem{Kohlhammer2011Solving}
J.~Kohlhammer, D.~Keim, M.~Pohl, G.~Santucci, and G.~Andrienko.
\newblock Solving problems with visual analytics.
\newblock {\em Procedia Computer Science}, 7:117--120, 2011.

\bibitem{Kwon12016Comparative}
B.~C. Kwon and B.~Lee.
\newblock A comparative evaluation on online learning approaches using parallel
  coordinate visualization.
\newblock In {\em Proceedings of SIGCHI Conference on Human Factors in
  Computing Systems}, pages 993--997, 2016.

\bibitem{Lai2020Automatic}
C.~Lai, Z.~Lin, R.~Jiang, Y.~Han, C.~Liu, and X.~Yuan.
\newblock Automatic annotation synchronizing with textual description for
  visualization.
\newblock In {\em Proceedings of the 2020 SIGCHI Conference on Human Factors in
  Computing Systems}, pages 1--13, 2020.

\bibitem{Bongshin2006Task}
B.~Lee, C.~Plaisant, C.~S. Parr, J.-D. Fekete, and N.~Henry.
\newblock Task taxonomy for graph visualization.
\newblock In {\em Proceedings of the 2006 AVI Workshop on BEyond Time and
  Errors: Novel Evaluation Methods for Information Visualization}, pages 1--5,
  2006.

\bibitem{Li2023WhyAI}
H.~Li, Y.~Wang, Q.~V. Liao, and H.~Qu.
\newblock Why is {AI} not a panacea for data workers? an interview study on
  human-ai collaboration in data storytelling.
\newblock {\em arXiv.2304.08366}, 2023.

\bibitem{Li2023Notable}
H.~Li, L.~Ying, H.~Zhang, Y.~Wu, H.~Qu, and Y.~Wang.
\newblock Notable: On-the-fly assistant for data storytelling in computational
  notebooks.
\newblock In {\em Proceedings of SIGCHI Conference on Human Factors in
  Computing Systems}, 2023.

\bibitem{Li2023GeoCamera}
W.~Li, Z.~Wang, Y.~Wang, D.~Weng, L.~Xie, S.~Chen, H.~Zhang, and H.~Qu.
\newblock {GeoCamera}: Telling stories in geographic visualizations with camera
  movements.
\newblock In {\em Proceedings of the 2023 CHI Conference on Human Factors in
  Computing Systems}, 2023.

\bibitem{Li2023Diverse}
Y.~Li, Y.~Qi, Y.~Shi, Q.~Chen, N.~Cao, and S.~Chen.
\newblock Diverse interaction recommendation for public users exploring
  multi-view visualization using deep learning.
\newblock {\em IEEE Transactions on Visualization and Computer Graphics},
  29(1):95--105, 2023.

\bibitem{Lin2023DMiner}
Y.~Lin, H.~Li, A.~Wu, Y.~Wang, and H.~Qu.
\newblock {DMiner}: Dashboard design mining and recommendation.
\newblock {\em IEEE Transactions on Visualization and Computer Graphics}, pages
  1--15, 2023.

\bibitem{Liu2023Evaluating}
C.~Liu and B.~Wu.
\newblock Evaluating large language models on graphs: Performance insights and
  comparative analysis.
\newblock {\em arXiv.2308.11224}, 2023.

\bibitem{Liu2020AutoCaption}
C.~Liu, L.~Xie, Y.~Han, D.~Wei, and X.~Yuan.
\newblock {AutoCaption}: An approach to generate natural language description
  from visualization automatically.
\newblock In {\em Proceedings of IEEE Pacific Visualization Symposium}, pages
  191--195, 2020.

\bibitem{Liu2020Paths}
Y.~Liu, T.~Althoff, and J.~Heer.
\newblock {Paths Explored, Paths Omitted, Paths Obscured}: Decision points \&
  selective reporting in end-to-end data analysis.
\newblock In {\em Proceedings of the SIGCHI Conference on Human Factors in
  Computing Systems}, pages 1--14, 2020.

\bibitem{Liu2022Fill}
Z.~Liu, C.~Chen, J.~Wang, X.~Che, Y.~Huang, J.~Hu, and Q.~Wang.
\newblock {Fill in the Blank}: Context-aware automated text input generation
  for mobile gui testing.
\newblock In {\em Proceedings of International Conference on Software
  Engineering}, pages 1355--1367, 2023.

\bibitem{Luo2022Natural}
Y.~Luo, N.~Tang, G.~Li, J.~Tang, C.~Chai, and X.~Qin.
\newblock Natural language to visualization by neural machine translation.
\newblock {\em IEEE Transactions on Visualization and Computer Graphics},
  28(1):217--226, 2022.

\bibitem{Maddigan2023Chat2VIS}
P.~Maddigan and T.~Susnjak.
\newblock {Chat2VIS}: Generating data visualizations via natural language using
  chatgpt, codex and gpt-3 large language models.
\newblock {\em IEEE Access}, 11:45181--45193, 2023.

\bibitem{Moritz2019Formalizing}
D.~Moritz, C.~Wang, G.~L. Nelson, H.~Lin, A.~M. Smith, B.~Howe, and J.~Heer.
\newblock Formalizing visualization design knowledge as constraints: Actionable
  and extensible models in draco.
\newblock {\em IEEE Transactions on Visualization and Computer Graphics},
  25(1):438--448, 2019.

\bibitem{Narechania2021NL4DV}
A.~Narechania, A.~Srinivasan, and J.~Stasko.
\newblock {NL4DV}: A toolkit for generating analytic specifications for data
  visualization from natural language queries.
\newblock {\em IEEE Transactions on Visualization and Computer Graphics},
  27(2):369--379, 2021.

\bibitem{Openai2023GPT4}
OpenAI.
\newblock {GPT-4} technical report.
\newblock {\em arxiv.2303.08774}, 2023.

\bibitem{Peng2021Mixed}
L.~Peng, Y.~Zhao, Y.~Hou, Q.~Wang, S.~Shen, X.~Lai, J.~Gao, J.~Dong, Z.~Lin,
  and S.~Chen.
\newblock Mixed-initiative visual exploration of social media text and events.
\newblock In {\em Proceedings of the IEEE Conference on Visualization and
  Visual Analytics}, 2021.

\bibitem{Perez2022Typology}
I.~P{\'{e}}rez-Messina, D.~Ceneda, M.~El-Assady, S.~Miksch, and F.~Sperrle.
\newblock A typology of guidance tasks in mixed-initiative visual analytics
  environments.
\newblock {\em Computer Graphics Forum}, 41(3):465--476, 2022.

\bibitem{Qian2021Learning}
X.~Qian, R.~A. Rossi, F.~Du, S.~Kim, E.~Koh, S.~Malik, T.~Y. Lee, and J.~Chan.
\newblock Learning to recommend visualizations from data.
\newblock In {\em Proceedings of the ACM SIGKDD Conference on Knowledge
  Discovery \& Data Mining}, pages 1359--1369, 2021.

\bibitem{Qin2023Toolllm}
Y.~Qin, S.~Liang, Y.~Ye, K.~Zhu, L.~Yan, Y.~Lu, Y.~Lin, X.~Cong, X.~Tang,
  B.~Qian, S.~Zhao, R.~Tian, R.~Xie, J.~Zhou, M.~Gerstein, D.~Li, Z.~Liu, and
  M.~Sun.
\newblock {ToolLLM}: Facilitating large language models to master 16000+
  real-world apis.
\newblock {\em arxiv.2307.16789}, 2023.

\bibitem{Sevastjanova2021Visinreport}
R.~Sevastjanova, M.~El-Assady, A.~Bradley, C.~Collins, M.~Butt, and D.~Keim.
\newblock {VisInReport}: Complementing visual discourse analytics through
  personalized insight reports.
\newblock {\em IEEE Transactions on Visualization and Computer Graphics},
  28(12):4757--4769, 2022.

\bibitem{Shen2022Towards}
L.~Shen, E.~Shen, Y.~Luo, X.~Yang, X.~Hu, X.~Zhang, Z.~Tai, and J.~Wang.
\newblock Towards natural language interfaces for data visualization: A survey.
\newblock {\em IEEE Transactions on Visualization and Computer Graphics},
  29(6):3121--3144, 2022.

\bibitem{Shi2021Autoclips}
D.~Shi, F.~Sun, X.~Xu, X.~Lan, D.~Gotz, and N.~Cao.
\newblock {AutoClips}: An automatic approach to video generation from data
  facts.
\newblock In {\em Computer Graphics Forum}, volume~40, pages 495--505. Wiley
  Online Library, 2021.

\bibitem{Sperrle2022Lotse}
F.~Sperrle, D.~Ceneda, and M.~El-Assady.
\newblock Lotse: A practical framework for guidance in visual analytics.
\newblock {\em IEEE Transactions on Visualization and Computer Graphics},
  29(1):1124--1134, 2022.

\bibitem{Stasko2005Information}
J.~Stasko, R.~Amar, and J.~Eagan.
\newblock Low-level components of analytic activity in information
  visualization.
\newblock In {\em Proceedings of the 2005 IEEE Symposium on Information
  Visualization}, page~15, 2005.

\bibitem{Stoiber2022Perspectives}
C.~Stoiber, D.~Ceneda, M.~Wagner, V.~Schetinger, T.~Gschwandtner, M.~Streit,
  S.~Miksch, and W.~Aigner.
\newblock Perspectives of visualization onboarding and guidance in va.
\newblock {\em Visual Informatics}, 6(1):68--83, 2022.

\bibitem{Christina2019Visualization}
C.~Stoiber, F.~Stoiber, M.~Pohl, H.~Stitz, M.~Streit, and W.~Aigner.
\newblock Visualization onboarding: Learning how to read and use
  visualizations.
\newblock In {\em Proceedings of VisComm Workshop at IEEE VIS Conference},
  2019.

\bibitem{sullivan2012using}
G.~M. Sullivan and R.~Feinn.
\newblock Using effect size—or why the {P} value is not enough.
\newblock {\em Journal of graduate medical education}, 4(3):279--282, 2012.

\bibitem{Tableau2003Embedding}
Tableau.
\newblock Tableau embedding api.
\newblock
  \url{https://help.tableau.com/current/api/embedding_api/en-us/index.html},
  2003.

\bibitem{Tamkin2021Understanding}
A.~Tamkin, M.~Brundage, J.~Clark, and D.~Ganguli.
\newblock Understanding the capabilities, limitations, and societal impact of
  large language models.
\newblock {\em arxiv.2102.02503}, 2021.

\bibitem{Tanahashi2016Study}
Y.~Tanahashi, N.~Leaf, and K.-L. Ma.
\newblock A study on designing effective introductory materials for information
  visualization.
\newblock {\em Computer Graphics Forum}, 35(7):117--126, 2016.

\bibitem{Thomas2005Illuminating}
J.~J. Thomas and K.~A. Cook.
\newblock {\em Illuminating the Path: An R\&D Agenda for Visual Analytics},
  pages 69--104.
\newblock IEEE Press, 2005.

\bibitem{Wang2022Self}
Y.~Wang, Y.~Kordi, S.~Mishra, A.~Liu, N.~A. Smith, D.~Khashabi, and
  H.~Hajishirzi.
\newblock {Self-Instruct}: Aligning language model with self generated
  instructions.
\newblock {\em 2212.10560}, 2022.

\bibitem{Wang2020DataShot}
Y.~Wang, Z.~Sun, H.~Zhang, W.~Cui, K.~Xu, X.~Ma, and D.~Zhang.
\newblock {DataShot}: Automatic generation of fact sheets from tabular data.
\newblock {\em IEEE Transactions on Visualization and Computer Graphics},
  26(1):895--905, 2020.

\bibitem{Wu2023Defence}
A.~Wu, D.~Deng, F.~Cheng, Y.~Wu, S.~Liu, and H.~Qu.
\newblock In defence of visual analytics systems: Replies to critics.
\newblock {\em IEEE Transactions on Visualization and Computer Graphics},
  29(1):1026--1036, 2023.

\bibitem{Yalcin2016Systematic}
M.~A. Yalcin.
\newblock {\em A systematic and minimalist approach to lower barriers in visual
  data exploration}.
\newblock PhD thesis, University of Maryland, College Park, 2016.

\bibitem{Yousefi2023Incontext}
S.~Yousefi, L.~Betthauser, H.~Hasanbeig, A.~Saran, R.~Millière, and
  I.~Momennejad.
\newblock {In-Context} learning in large language models: A
  neuroscience-inspired analysis of representations.
\newblock {\em arxiv.2310.00313}, 2023.

\bibitem{Zha2023Tablegpt}
L.~Zha, J.~Zhou, L.~Li, R.~Wang, Q.~Huang, S.~Yang, J.~Yuan, C.~Su, X.~Li,
  A.~Su, T.~Zhang, C.~Zhou, K.~Shou, M.~Wang, W.~Zhu, G.~Lu, C.~Ye, Y.~Ye,
  W.~Ye, Y.~Zhang, X.~Deng, J.~Xu, H.~Wang, G.~Chen, and J.~Zhao.
\newblock {TableGPT}: Towards unifying tables, nature language and commands
  into one {GPT}.
\newblock {\em arxiv.2307.08674}, 2023.

\bibitem{Zhang2023Data}
W.~Zhang, Y.~Shen, W.~Lu, and Y.~Zhuang.
\newblock {Data-Copilot}: Bridging billions of data and humans with autonomous
  workflow.
\newblock {\em arxiv.2306.07209}, 2023.

\bibitem{Zhao2021Chartstory}
J.~Zhao, S.~Xu, S.~Chandrasegaran, C.~Bryan, F.~Du, A.~Mishra, X.~Qian, Y.~Li,
  and K.-L. Ma.
\newblock Chartstory: Automated partitioning, layout, and captioning of charts
  into comic-style narratives.
\newblock {\em IEEE Transactions on Visualization and Computer Graphics},
  29(2):1384--1399, 2021.

\end{thebibliography}

\fi

\ifCLASSOPTIONcaptionsoff
  \newpage
\fi

\ifx\hideappendix\undefined
  \appendix


\section{System Specification}
\label{appendix:spec}

The following section outlines the detailed specifications for the two visual analytics systems we used. These specifications could be examples for LLMs to generate tutorials for a new VA system.

\subsection{VAST Challenge System}
\label{appendix:spec-vast}

We provide the specification of the VAST challenge system we used in Sec.6.1.

\begin{lstlisting}[language=json]
{
  "systemInfo": {
    "systemName": "Social media event analysis system",
    "viewNumber": 5
  },
  "viewInfo": {
    "viewStyleInfo": [
      {
        "viewName": "Timeline",
        "layers": [
          {
            "markType": "area",
            "encoding": {
              "x": {"field": "datetime", "type": "temporal"},
              "y": [
                {"field": "message count", "type": "quantitative", "description": "microblog data"},
                {"field": "ccdata count", "type": "quantitative", "description": "call center data"}
              ],
              "color": {"field": "message type", "type": "nominal"}
            }
          }
        ]
      },
      {
        "viewName": "Risk Map",
        "layers": [
          {
            "markType": "hexagon",
            "encoding": {
              "x": {"field": "Latitude", "type": "quantitative"},
              "y": {"field": "Longitude", "type": "quantitative"},
              "lod": {"field": "Risk Level", "type": "quantitative", "description": "Risk Level is positively correlated with the number of messages and negative emotions."
              "color": {"field": "Risk Level", "type": "quantitative"},
              }
            },
            "tooltip": [
              {"field": "Risk Level", "type": "quantitative"},
              {"field": "Messages", "type": "text"}
            ]
          },
          {
            "markType": "square",
            "encoding": {
              "x": {"field": "Latitude", "type": "quantitative"},
              "y": {"field": "Longitude", "type": "quantitative"},
              "color": {"field": "Importance Level", "type": "quantitative"}
            },
            "tooltip": [
              {"field": "Risk Level", "type": "quantitative"},
              {"field": "Important Level", "type": "quantitative"},
              {"field": "Messages", "type": "text"},
            ]
          },
          {
            "markType": "point",
            "encoding": {
              "x": [{"field": "Latitude", "type": "quantitative"},{"field": "Latitude", "type": "quantitative"}],
              "y": [{"field": "Longitude", "type": "quantitative"},{"field": "Longitude", "type": "quantitative"}],
              "color": {"field": "message type", "type": "nominal"}
            }
          },
        ]
      },
      {
        "viewName": "Graph",
        "layers": [
          {
            "markType": "circle",
            "encoding": {
              "nodes": {"field": "node ID", "type": "nominal"},
              "color": {"field": "group", "type": "nominal"},
              "size": {"field": "degree", "type": "quantitative"}
            }
          },
          {
            "markType": "line",
            "encoding": {
              "source": {"field": "node ID", "type": "nominal"},
              "target": {"field": "node ID", "type": "nominal"},
              "size": {"field": "co-occurrence", "type": "quantitative"}
            }
          }
        ]
      },
      {
        "viewName": "Message",
        "layers": [
          {
            "markType": "card",
            "encoding": {
              "title": {"field": "time", "type": "temporal"},
              "content": {"field": "message", "type": "text"}
            }
          }
        ]
      },
      {
        "viewName": "Keyword",
        "layers": [
          {
            "markType": "text",
            "encoding": {
              "size": {"field": "frequency", "type": "quantitative"},
              "color": {"field": "frequency", "type": "quantitative"}
            }
          }
        ]
      }
    ]
  },
  "viewCoordinationInfo": [
    {
      "sourceViewName": "Timeline",
      "targetViewName": ["Risk Map", "Graph", "Messages", "Keyword"],
      "coordinationType": "filter",
      "interaction": {"type": "click"}
    },
    {
      "sourceViewName": "Keyword",
      "targetViewName": ["Timeline"],
      "coordinationType": "filter",
      "interaction": {
        "type": "click",
        "effect": {
          "action": "addCategory",
          "targetView": "Timeline",
          "category": "event",
          "changeby": "dropdown"
        }
      }
    }
  ]
}
\end{lstlisting}

\subsection{Tableau Superstore System}


We also provide the specification of the tableau's system we used in Sec.6.2.

\begin{lstlisting}[language=json]
{
  "systemInfo": {
    "systemName": "Superstore system",
    "viewNumber": 9
  },
  "viewsInfo": {
    "viewStyleInfo": [
      {
        "viewName": "Sales | By Category",
        "layers": [
          {
            "markType": "Bar",
            "encoding": {
              "x": [
                {"field": "AGG(Metric Swap Calc (CY))", "type": "quantitative"},
                {"field": "AGG(Metric Swap Calc (PY))", "type": "quantitative"}
              ],
              "y": {"field": "Category", "type": "nominal"}
            }
          }
        ]
      },
      {
        "viewName": "Sales | By Manufacturer",
        "layers": [
          {
            "markType": "Bar",
            "encoding": {
              "x": {"field": "Manufacturer", "type": "nominal"},
              "y": {"field": "AGG(Metric Swap Calc (CY))", "type": "quantitative"}
            }
          }
        ]
      },
      {
        "viewName": "Sales | By Segment",
        "layers": [
          {
            "markType": "Bar",
            "encoding": {
              "x": [
                {"field": "AGG(Metric Swap Calc (CY))", "type": "quantitative"},
                {"field": "AGG(Metric Swap Calc (PY))", "type": "quantitative"}
              ],
              "y": {"field": "Segment", "type": "nominal"}
            }
          }
        ]
      },
      {
        "viewName": "Sales | By State",
        "layers": [
          {
            "markType": "circle",
            "encoding": {
              "x": {"field": "AGG(Metric Swap Calc (CY))", "type": "quantitative"},
              "y": {"field": "State/Province", "type": "nominal"},
              "size": {"field": "AGG(Metric Swap Calc (CY))", "type": "quantitative"}
            }
          }
        ]
      },
      {
        "viewName": "Sales | By Sub-Category",
        "layers": [
          {
            "markType": "Bar",
            "encoding": {
              "x": {"field": "Sub-Category", "type": "nominal"},
              "y": [
                {"field": "AGG(Metric Swap Calc (CY))", "type": "quantitative"},
                {"field": "AGG(Metric Swap Calc (PY))", "type": "quantitative"}
              ]
            }
          }
        ]
      },
      {
        "viewName": "Profit KPI (Line)",
        "layers": [
          {
            "markType": "Line",
            "encoding": {
              "x": {"field": "Profit", "type": "quantitative"},
              "y": {"field": "MONTH(Order Date)", "type": "temporal"}
            }
          }
        ]
      },
      {
        "viewName": "Sales KPI (Line)",
        "layers": [
          {
            "markType": "Line",
            "encoding": {
              "x": {"field": "Sales", "type": "quantitative"},
              "y": {"field": "MONTH(Order Date)", "type": "temporal"}
            }
          }
        ]
      },
      {
        "viewName": "Total Orders KPI (Area)",
        "layers": [
          {
            "markType": "Automatic",
            "encoding": {
              "x": {"field": "CNTD(Order ID)", "type": "quantitative"},
              "y": {"field": "MONTH(Order Date)", "type": "temporal"}
            }
          }
        ]
      },
      {
        "viewName": "Total Customers KPI (Area)",
        "layers": [
          {
            "markType": "Automatic",
            "encoding": {
              "x": {"field": "CNTD(Customer Name)", "type": "quantitative"},
              "y": {"field": "MONTH(Order Date)", "type": "temporal"}
            }
          }
        ]
      }
    ]
  },
  "viewCoordinationInfo": [
    {
      "sourceViewName": "Sales | By State",
      "targetViewName": [
        "Sales | By Manufacturer",
        "Sales | By Segment",
        "Sales | By Sub-Category",
        "Sales | By Category",
        "Profit KPI (Line)",
        "Sales KPI (Line)",
        "Total Orders KPI (Area)",
        "Total Customers KPI (Area)"
      ],
      "coordinationType": "filter",
      "interaction": {"type": "click"}
    },
    {
      "sourceViewName": "Sales | By Segment",
      "targetViewName": [
        "Sales | By Manufacturer",
        "Sales | By Sub-Category",
        "Sales | By Category",
        "Sales | By State",
        "Profit KPI (Line)",
        "Sales KPI (Line)",
        "Total Orders KPI (Area)",
        "Total Customers KPI (Area)"
      ],
      "coordinationType": "filter",
      "interaction": {"type": "click"}
    },
    {
      "sourceViewName": "Sales | By Category",
      "targetViewName": [
        "Sales | By Manufacturer",
        "Sales | By Segment",
        "Sales | By Category",
        "Sales | By State",
        "Profit KPI (Line)",
        "Sales KPI (Line)",
        "Total Orders KPI (Area)",
        "Total Customers KPI (Area)"
      ],
      "coordinationType": "filter",
      "interaction": {"type": "click"}
    },
    {
      "sourceViewName": "Sales | By Sub-Category",
      "targetViewName": [
        "Sales | By Manufacturer",
        "Sales | By Segment",
        "Sales | By Category",
        "Sales | By State",
        "Profit KPI (Line)",
        "Sales KPI (Line)",
        "Total Orders KPI (Area)",
        "Total Customers KPI (Area)"
      ],
      "coordinationType": "filter",
      "interaction": {"type": "click"}
    },
    {
      "sourceViewName": "Sales | By Manufacturer",
      "targetViewName": [
        "Sales | By Segment",
        "Sales | By Sub-Category",
        "Sales | By Category",
        "Sales | By State",
        "Profit KPI (Line)",
        "Sales KPI (Line)",
        "Total Orders KPI (Area)",
        "Total Customers KPI (Area)"
      ],
      "coordinationType": "filter",
      "interaction": {"type": "click"}
    }
  ]
}
\end{lstlisting}

\label{appendix:spec-tableau}

\section{Prompt Examples}
\label{appendix:prompt}
Within this appendix, we showcase several exemplary prompts encompassing various functionalities of our framework. These prompts elucidate how the framework can be utilized across different stages of the visual analytics process.

\subsection{Onbaording}
The following prompt aims at generating an onboarding tutorial based on the specifications of a given system:

\begin{coloredquotation}
    \textbf{Prompt example for onboarding:}\\
    Here are the specifications of a visual analytics system. \\
    (Add a specification)\\
    The specification includes the system-level, view-level and views’ coordination information. You need to introduce each view with style (data meaning, visual mapping) and the relationship between views. Please give your answer in the following JSON format:
\begin{lstlisting}[style=mystyle]
{"viewName": "", 
 "content": 
- <b>Style</b>: ""<br>
- <b>Coordination<b>: ""<br>
}
\end{lstlisting}
\end{coloredquotation}

An output example of a tutorial:

\begin{coloredquotationoutput}
\begin{lstlisting}[style=mystyle]
[
  {
    "viewName": "Sales | By Category",
    "content": {
    "viewName": "Sales | By Category",
    "content": 
      "<b>Style</b>: Displays sales data by category, using bar marks. The X-axis represents both the current year and previous year metrics, while the Y-axis represents the categories.<br>",
      "<b>Coordination</b>: When a bar (representing a category) is clicked, it filters data in the views: 'Sales | By Manufacturer', 'Sales | By Segment', 'Sales | By Sub-Category', 'Sales | By State', 'Profit KPI (Line)', 'Sales KPI (Line)', 'Total Orders KPI (Area)', 'Total Customers KPI (Area)'.<br>"
  },
  }, ...
]
\end{lstlisting}
\end{coloredquotationoutput}

\subsection{Recommending Insights}
In this section, we outline the prompt examples that serve as a flexible way to instruct insight recommendation tasks. 

The first template focuses on recommending insights types based on the user's selections and choosing functions to calculate insights. Here is an example prompt:

\begin{coloredquotation}
\textbf{Prompt example for insight calculation:}\\
When the user makes an action on a view, the system changes. 
\begin{lstlisting}[style=mystyle]
Current selections: 
[{"dimName": "Segment", "value": "Consumer","viewName": "Sales | By Segment"},
{"dimName": "State/Province", "value": "California","viewName": "Sales | By State"},
{"dimName": "State/Province", "value": "New York","viewName": "Sales | By State"}]
}
\end{lstlisting}
\begin{lstlisting}[style=mystyle]
viewStyleInfo
{
  "viewName": "Sales | By Category",
  "markType": "Bar",
  "encoding": {
    "x": [
      "Metric Swap Calc (% Chg)(Label)",
      "AGG(Metric Swap Calc (CY))",
      "AGG(Metric Swap Calc (PY))"
    ],
    "y": [
      "Category"
      ]
  }, ...
Coordination information
{
  "sourceViewName": "Sales | By Segment",
  "targetViewName": [
    "Sales | By Manufacturer",
    "Sales | By Sub-Category",
    "Sales | By Category",
    "Sales | By State",
    "Profit KPI (Line)",
    "Sales KPI (Line)",
    "Total Orders KPI (Area)",
    "Total Customers KPI (Area)"
  ],
  "coordinationType": "filter",
  "interaction": {
    "type": "click"
  }
},
\end{lstlisting}

According to the data info in each view and the analytical task, you should select all suitable analytical functions related to the user’s task. If a view contains more than one measure, all measures need to be considered to find appropriate functions. Cross-view insights, such as correlation, are allowed. You also need to give a relevance score to assess how closely related the insight type is to the task.\\
\begin{lstlisting}[style=mystyle]
{"user task": "Analyze the sales of the superstore from different perspectives."}
"functions": [
  {
    "name": "get_change_point",
    "description": "Get the change point in a time series dataset"
  },{
    "name": "get_seasonality",
    "description": "Get the seasonality in a time series dataset"
  },{
    "name": "get_trend",
    "description": "Get the trend in a time series dataset"
  },{
    "name": "get_outlier",
    "description": "Get outliers in a dataset",
  },{
    "name": "get_correlation",
    "description": "Get the correlation of two time series datasets"
  },{
    "name": "get_outstanding_top1",
    "description": "Get the first n most values that are significantly larger than other values in the data column",
  },{
    "name": "get_outstanding_last",
    "description": "Get the last item in a data column that is significantly smaller than the other values"
  }
]
\end{lstlisting}
For all possible insight types, you need to output a JSON list containing view name, function name, variables name, and relevanceScore. The viewName and variableName can be listed if they have two variables on two views, such as calculating correlation. Your output should be like this:
\begin{lstlisting}[style=mystyle]
[{"viewName":"", "functionName":"", "variableName":"", "relevanceScore":}].
\end{lstlisting}
\end{coloredquotation}

After the functions return the results of insights with the significance, the next step is to evaluate the impact of these insights. This next template assists in converting the raw data results into more interpretable sentences and quantifies the potential impact of the insights on the user's tasks. Here's an example:

\begin{coloredquotation}
    \textbf{Prompt example for insight assessment:}\\
    The selected insights are implemented and return the result, including the value and significance score. The data is sorted by significance score. The dimName is the x-axis, value is the answer's x-axis value, variableName is the answer's column name (generally y-axis), result is the answer, and parameters used to describe the answer (e.g., 'Same' means two series have the same direction in correlation calculation. \\
    I need you to translate the answer into a sentence. The answer should contain more info to explain in detail. You also need to give an impact score. The impact score quantifies the potential business impact of the insight. Please give your answer in the following format:
\begin{lstlisting}[style=mystyle]
[{"impactScore":, "answer":""}]
\end{lstlisting}
\end{coloredquotation}

Lastly, the following template is designed for situations where there are multiple rounds of insight generation. Its purpose is to compare and identify the similarities or differences between the datasets of insights obtained in different rounds or conditions. Here's an example:

\begin{coloredquotation}
    \textbf{Prompt example for comparing insights:}\\
    From two rounds, identify the similarity or difference of the same type of insights in two insight datasets. If we do not select a data element, then the insights are aggregated values for all data, but if we select one or two elements, the insights are aggregated for those two elements. Please use a sentence to describe the similarity or difference. And do not use round. You should use the data name to represent the round. Your output format should be like this: 
\begin{lstlisting}[style=mystyle]
[{"answer":"", "viewName":"", "dimName":"", "value":""}]
\end{lstlisting}
\end{coloredquotation}

An output example of insights:
\begin{coloredquotationoutput}
\begin{lstlisting}[style=mystyle]
[{'question': '', 'viewName': 'Sales | By State', 'dimName': 'State/Province', 'value': ['California', 'New York'], 'functionName': 'get_outstanding_top2', 'variableName': 'AGG(Metric Swap Calc (CY))', 'result': [57682.45549999996, 36906.00399999999], 'significance_score': 1.0, 'parameters': None, 'relevance_score': 0.9, 'impact_score': 0.9, 'answer': 'The states with the highest sales in the current year are California and New York with sales amounting to 57682.46 and 36906.00 respectively.', 'final_score': 0.5},
{'question': '', 'viewName': 'Sales KPI (Line)', 'dimName': 'MONTH(Order Date)', 'value': '2022-03-01 00:00:00', 'functionName': 'get_change_point', 'variableName': 'SUM(Sales)', 'result': 526.9643807018581, 'significance_score': 0.9999999999285639, 'parameters': None, 'relevance_score': 0.9, 'impact_score': 0.9, 'answer': 'There was a significant change in sales in March 2022, with the sales amounting to 526.96.', 'final_score': 0.4999999999996},
{'question': '', 'viewName': 'Sales KPI (Line)', 'dimName': None, 'value': None, 'functionName': 'get_trend', 'variableName': 'SUM(Sales)', 'result': 'increase', 'significance_score': 0.9999999794746961, 'parameters': None, 'relevance_score': 0.9, 'impact_score': 0.9, 'answer': 'The sales has been increasing.', 'final_score': 0.4999999989737348},
{'question': '', 'viewName': 'Sales | By Manufacturer', 'dimName': 'Manufacturer', 'value': ['Other', 'Canon'], 'functionName': 'get_outstanding_top2', 'variableName': 'AGG(Metric Swap Calc (CY))', 'result': [36727.54599999997, 27459.864], 'significance_score': 0.9920215895038219, 'parameters': None, 'relevance_score': 0.9, 'impact_score': 0.9, 'answer': 'The manufacturers with the highest sales in the current year are Other and Canon with sales amounting to 36727.55 and 27459.86 respectively.', 'final_score': 0.4996010794751911},
...] 
\end{lstlisting}

\end{coloredquotationoutput}

\begin{table*}[ht]
    \caption{The table presents commonly used insight types in visual analytics.}
    \label{table:insight-type}
    \begin{tabularx}{\textwidth}{lX}
        \toprule
        \textbf{Insight type} & \textbf{Description} \\
        \midrule
        Outstanding No.1 & The leading value is significantly higher than all the remaining values. \\
        Outstanding Top 2 & The leading two values are significantly higher than all the remaining values.\\
        Outstanding Last & The value is remarkably lower than all the remaining values.\\
        Attribution & The leading value dominates (accounting for $\geq 50\%$) the group. \\
        Change Point & A specific point in time where there is a significant change or shift in the underlying data-generating process. \\
        Outlier & An observation or data point that significantly deviates from the rest of the data.\\
        Seasonality & A regular and predictable pattern of fluctuations or variations that occur at specific intervals of time.\\
        Trend & A time series has an obvious trend (increase or decrease) with a certain turbulence level (steadily/ with turbulence). \\
        Correlation & The statistical relationships between random variables, multidimensional data or time series.\\
        Difference & The similarity or difference between two or more datasets\\
        Aggregation & The descriptive statistical indicators (e.g., average, sum, count, etc.) based on the data attributes.\\
        Value & The the exact value of data attribute(s) under specific criteria.\\
        Text summary & The core ideas of a text dataset. The summary might have spatial or temporal features.\\
        Important nodes or links & The important nodes or links in a graph under specific criteria.\\
        Important text or keywords & The important original texts or keywords under specific criteria. \\
        \bottomrule
    \end{tabularx}
\end{table*}

\subsection{Generate Report}
In this section, we outline the prompt examples to generate reports.

The first template is designed to help users create insightful reports based on historical analysis data of the system. Here, for the first step, the user's interaction, we set the type ``click'' as type to let LLMs know this piece of data, while for the second step, we set the type using the function description for better clarification of this step's insight. The addition of type makes the LLMs more aware of the source of insight. 

\begin{coloredquotation}
    \textbf{Prompt template for report summarization:}\\
    Here is a historical analysis of the system data. The data includes user interactions and insights at each step of the analysis:\\
\begin{lstlisting}[style=mystyle]
[{"step": "step 1", "insights": {"type": "click", "value": ["19:35", "19:43"], "viewName": "Timeline"}, "image_name": "Timeline_1_0.png"},
{"step": "step 2", "insights": {"type": "The events in top 2 risk-level regions during a given time period", "value": [{"title": "Gelatogalore Shooting Incident", "label": "Shooting incident involving a black van and police at Gelatogalore leading to a cop's death."}, {"title": "Biker Henri Incident", "label": "Biker Henri was hit and consoled with coffee at a local shop."}], "viewName": "Risk Map"}, "imageName": "RiskMap_1_1.png"},...]
\end{lstlisting}
    Your task is to write an insight report to present these findings. The amount of insight should be equal to the number of steps for given data. Ensure you include both a cover(report title) and a conclusion. 
\end{coloredquotation}

The following template aids users in converting their summarized reports into LaTeX-formatted slides. 

\begin{coloredquotation}
    \textbf{Prompt template for LaTeX code generation:}\\
    Transform the summarized report into LaTeX slides. For each slide, if an insight exhibits a clear hierarchy, segment it using bullet points. Accompany each insight with a screenshot from the system. The image name for these screenshots can be found in the historical analysis data. Please integrate the following commands for style configuration.\\
\begin{lstlisting}[style=mystyle]
\documentclass{beamer}
\usetheme{Madrid}
\usecolortheme{seahorse}
\includegraphics[height=0.6\textheight,keepaspectratio]{../image/image_name.png}.
\end{lstlisting}
    The image\_name is provided in the historical analysis data. 
\end{coloredquotation}

An output example of a report:
\begin{coloredquotationoutput}
\begin{lstlisting}[style=mystyle]
\documentclass{beamer}
\usetheme{Madrid}
\usecolortheme{seahorse}

\begin{document}
\title{Risk Event Analysis for Abila City}
\date{January 23, 2014}
\frame{\titlepage}

\begin{frame}{Specific Time Period of Interest}
\begin{itemize}
    \item During the evening, a specific time period between 19:34 to 19:43 surfaced as particularly significant.
\end{itemize}
\centering
\includegraphics[width=0.8\textwidth]{../image/Timeline_1_0.png}
\end{frame}

\begin{frame}{Major Risk Events}
\begin{itemize}
    \item Shooting incident at Gelatogalore
    \begin{itemize}
        \item Confrontation between a black van and the police.
        \item Resulted in a police officer's death.
    \end{itemize}
    \item Vehicular incident involving Biker Henri
    \begin{itemize}
        \item Biker Henri was hit.
        \item Later consoled with a coffee at a local shop.
    \end{itemize}
\end{itemize}
\centering
\includegraphics[width=0.6\textwidth]{../image/RiskMap_1_1.png}
\end{frame}

\begin{frame}{Messages Analysis}
\begin{itemize}
    \item Real-time reactions from the public on the incidents.
    \item Considerable sympathy and concern for Biker Henri.
    \item Chaos and fear during the Gelatogalore Shooting Incident.
\end{itemize}
\centering
\includegraphics[width=0.3\textwidth]{../image/Message_1_2.png}
\end{frame}

\begin{frame}{Graph-based Representation}
\begin{itemize}
    \item Nodes associated with two primary incidents.
    \item Hashtags and entities linked to each event.
\end{itemize}
\centering
\includegraphics[width=0.8\textwidth]{../image/Graph_1_3.png}
\end{frame}

\begin{frame}{Keyword Analysis}
\begin{itemize}
    \item Keywords associated with Gelatogalore Shooting Incident:
    \begin{itemize}
        \item 'shot', 'police', 'gelato', 'gelatogalore', 'van', 'cop'
    \end{itemize}
    \item Keywords linked with Biker Henri Incident:
    \begin{itemize}
        \item 'hit', 'run', 'coffee', 'park', 'driver'
    \end{itemize}
\end{itemize}
\centering
\includegraphics[width=0.5\textwidth]{../image/Keyword_1_4.png}
\end{frame}

\begin{frame}{Significance of 'van'}
\begin{itemize}
    \item The keyword 'van' was isolated as significant.
\end{itemize}
\centering
\includegraphics[width=0.5\textwidth]{../image/Keyword_1_5.png}
\end{frame}

\begin{frame}{Timeline Analysis of 'van'}
\begin{itemize}
    \item Sequence of events linked to the keyword 'van'.
    \item Highlights the pivotal role of a black van in the evening's events.
\end{itemize}
\centering
\includegraphics[width=0.8\textwidth]{../image/Timeline_1_6.png}
\end{frame}

\begin{frame}{Conclusion}
\begin{itemize}
    \item The evening was fraught with tension, characterized by two major risk events.
    \item Data-driven insights derived from analyses highlight the sequence, nature, and public sentiment of the events.
    \item Provides a foundation for future investigations and response strategies.
\end{itemize}
\end{frame}

\end{document}
\end{lstlisting}

\end{coloredquotationoutput}

\section{Evaluation of LLM's Data Analysis Ability}
\label{appendix:test}

In this section, we test the ability of gpt-3.5-turbo to calculate answers to the common insight tasks. The insight types are in Table~\ref{table:insight-type}. This test mainly considers the following two factors. First, whether GPT can calculate the accurate answer of a given task (accuracy). Second, the number of data rows LLMs can handle.

Due to the first factor, we tested 50 times, each using different data for an insight task. The data used each time is randomly generated within a certain range according to the same format (same number of data rows). For the second factor, in an insights task, tests are performed on different numbers of data rows (each 50 times). For example, for a single point insight finding the maximum value, the data format for each test is an array, and the length of the array is a variable parameter. For each fixed array length, test 50 times using randomly generated data. Besides, to facilitate the automation of the program to extract the answers to the GPT's responses, we let GPT output in a required format like ``Use \{\} to include the result". The results of our tests are in Fig.~\ref{fig:test_results}.

\begin{figure}[ht]
    \centering
    \includegraphics[width=\linewidth]{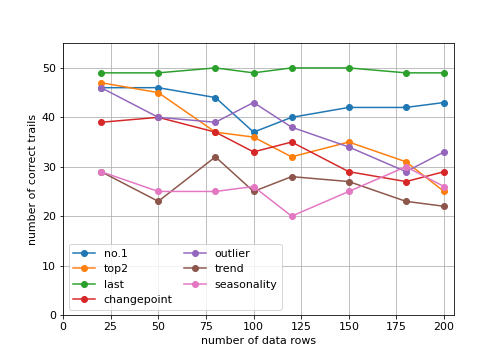}
    \caption{A test to assess the data analysis ability of LLMs. The y-axis represents the number of correct trials out of 50 trials. The x-axis shows the varying data rows, which include [20, 50, 80, 100, 120, 150, 180, 200]. The accuracy of most types of insight analysis tends to decrease as the rows of data increase.}
    \label{fig:test_results}
\end{figure}

Based on the test results, the accuracy of most types of insight analysis tends to decrease as the volume of data increases. 
Besides, there are notable differences in the performance of GPT across various analytical tasks. 

In certain singular shape insight analysis tasks, like identifying the ``outstanding last" or ``outstanding No. 1," GPT exhibits more favorable performance, and the decline in accuracy is less apparent with increasing data volume.
It is worth highlighting that with a large volume of data, the accuracy of the top 2 is notably lower than that of the No.1. 
In the false results produced by GPT, it is not uncommon for one of the top 2 predictions to be accurate while the other is incorrect. 
For example, in the 47th test ($number of data rows = 200$) of the ``outstanding top 2 values'' task, GPT output, ``The outstanding top 2 values are {998}, {993}." The correct top value is indeed 998, while the second largest number is 997, which differs from GPT's answer of 993.

Nonetheless, in time series-related tasks, such as ``trend" and ``seasonality" analysis, GPT performs inadequately. 
When determining the trend of a time series, GPT often makes mistakes when assessing data that lacks a distinct increasing or decreasing trend and instead exhibits fluctuations.
For example, in the 20th test ($number of data rows = 200$) of the ``trend" task, GPT output, ``The trend of the data in the time series over time is decreased. \{-1\}." However, there is no ascending or descending trend in this data.
In Fig.~\ref{fig:test_results}, it can be observed that their accuracy rate is consistently around 50\%, implying that the performance is nearly equivalent to random selection.

An example of a prompt during testing is as follows:

\begin{coloredquotation}
    \textbf{Prompt example for finding outstanding No.1 value:}\\
    I have a table that shows the values for each individual, each belonging to one category, for a total of three categories. 
\begin{lstlisting}[style=mystyle]
Data:
 category,individual_index,value
"category2",0,402
"category1",1,525
"category1",2,188
"category1",3,570
"category3",4,781
"category2",5,421
"category2",6,698
"category1",7,188
"category3",8,83
"category3",9,739

\end{lstlisting}
    My question is: What is the outstanding No.1 value for an individual? Use \{\} to include the result. For example, if the No.1 value is 970, please output \{970\} at the end.

\end{coloredquotation}

\fi



%


\ifx\hidebio\undefined
  \vspace{-25pt}
\begin{IEEEbiography}[{\includegraphics[width=1in,height=1.25in,clip,keepaspectratio]{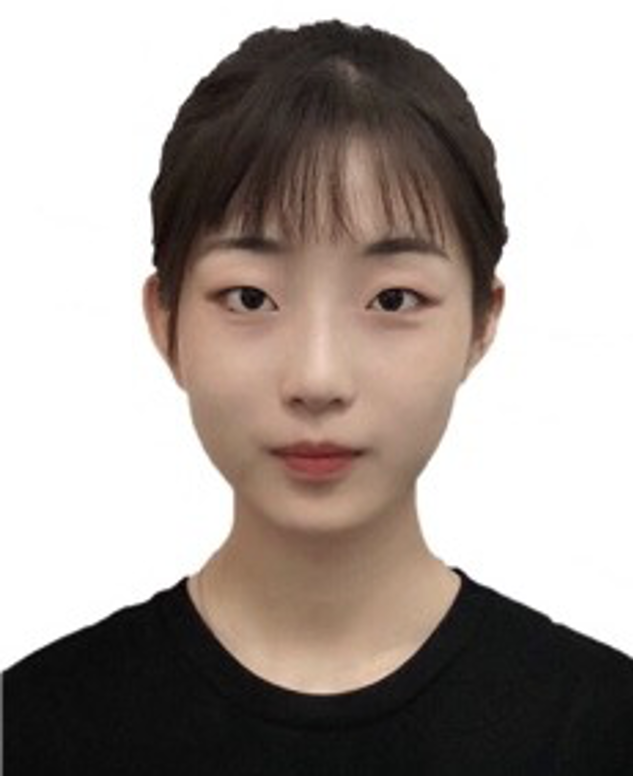}}]{Yuheng Zhao} is currently a Ph.D student at the School of Data Science, Fudan University. Her primary research interests are visualization and visual analytics, with an emphasis on foundation model-powered intelligent visual analytics. For more information, please visit https://https://www.yuhengzhao.me/.
\end{IEEEbiography}

\vspace{-25pt}

\begin{IEEEbiography}[{\includegraphics[width=1in,height=1.25in,clip,keepaspectratio]{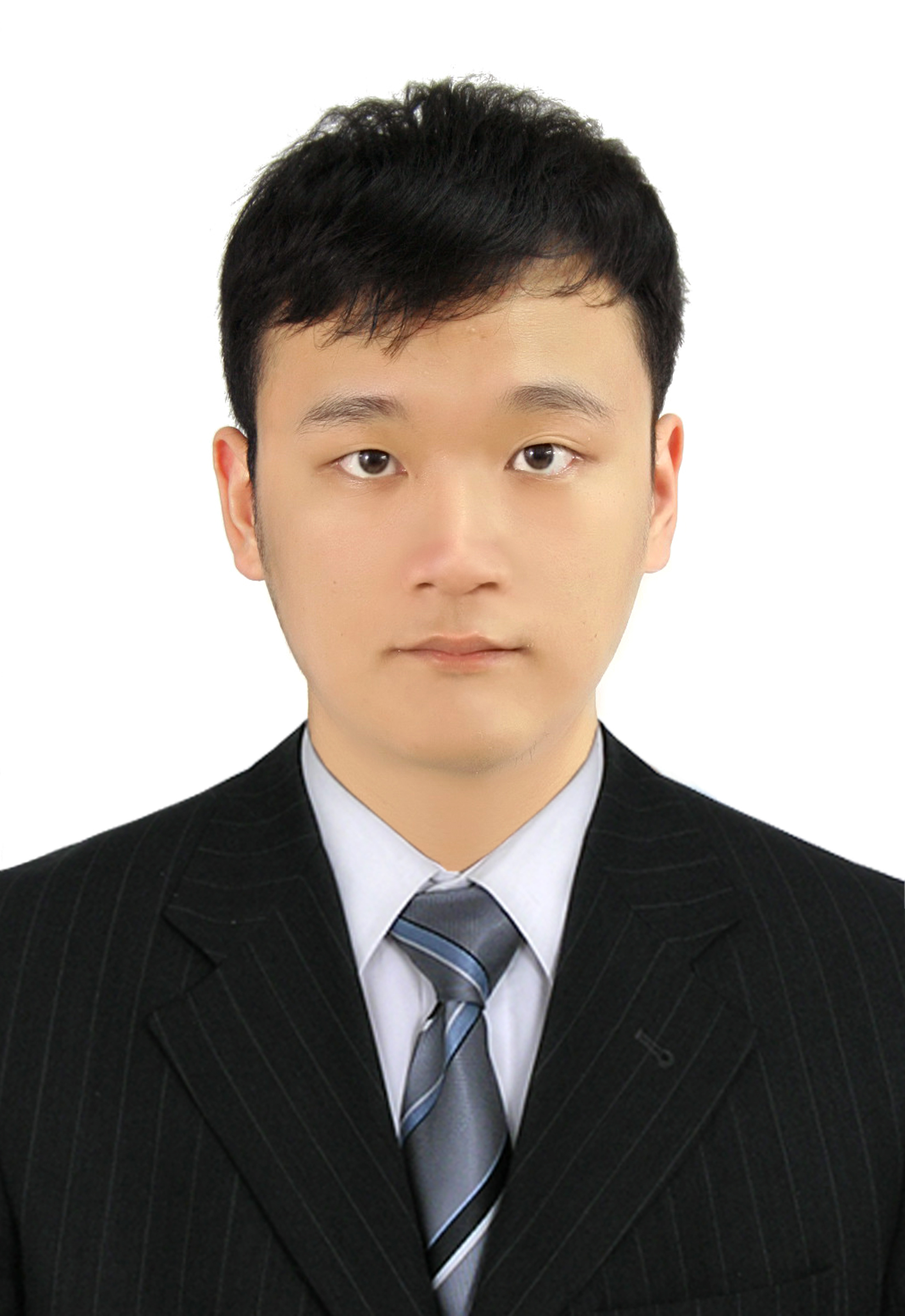}}]{Yixing Zhang}
received a B.S. in Computer Science and Technology from Hefei University of Technology. He is currently a Master's student at the School of Data Science, Fudan University. His primary research interests are visualization and visual analytics, with a specific focus on intelligent and adaptive visualization recommendation and generation.
\end{IEEEbiography}

\vspace{-25pt}

\begin{IEEEbiography}[{\includegraphics[width=1in,height=1.25in,clip,keepaspectratio]{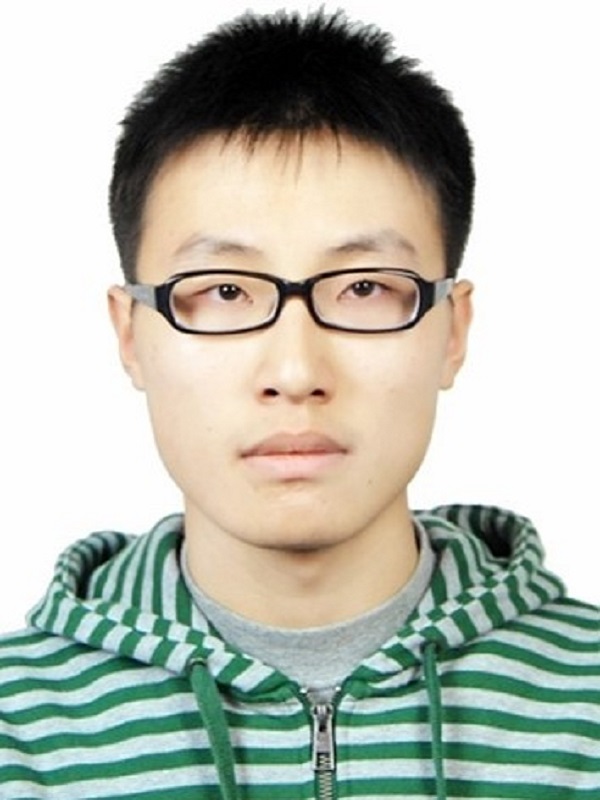}}]{Yu Zhang}
received a B.S. degree in Intelligence Science and Technology from Peking University in 2017. Since then, he has been pursuing a Ph.D. at the Department of Computer Science, University of Oxford. His research focuses on intelligent user interfaces in the field of human-computer interaction.
\end{IEEEbiography}

\vspace{-25pt}

\begin{IEEEbiography}[{\includegraphics[width=1in,height=1.25in,clip,keepaspectratio]{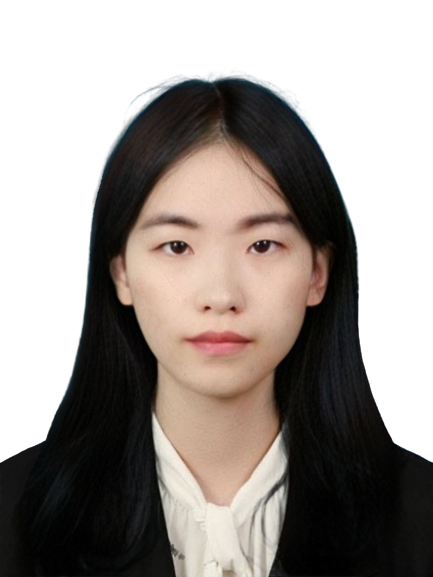}}]{Xinyi Zhao}
having earned a B.S. in Data Science and Big Data Technology from Fudan University, she is currently advancing her academic journey by pursuing a Master's degree in Data Science at Columbia University. 
Her primary research interests lie in applied machine learning, with a specific focus on enhancing data visualization techniques.
\end{IEEEbiography}

\vspace{-25pt}

\begin{IEEEbiography}[{\includegraphics[width=1in,height=1.25in,clip,keepaspectratio]{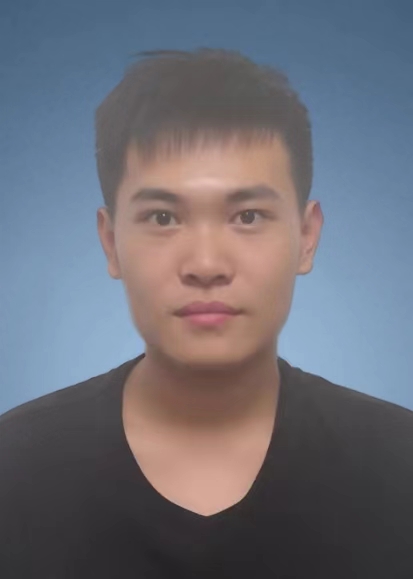}}]{Junjie Wang}
is currently a master's student at the School of Data Science, Fudan University. His primary research interests are visualization and visual analytics, with a specific focus on intelligent and adaptive visual analytics.
\end{IEEEbiography}

\vspace{-25pt}

\begin{IEEEbiography}[{\includegraphics[width=1in,height=1.25in,clip,keepaspectratio]{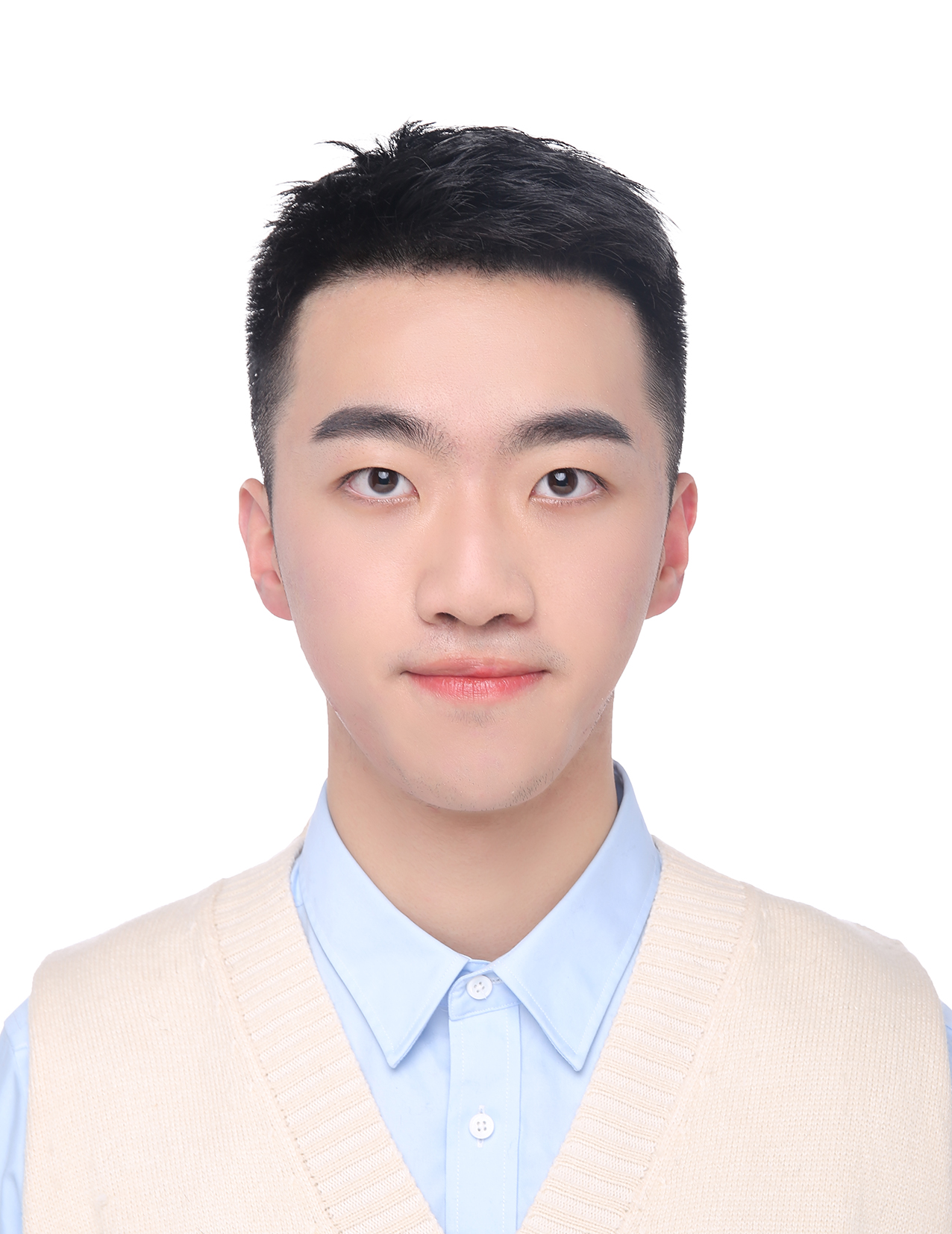}}]{Zekai Shao}
is a Ph.D. student at the School of Data Science, Fudan University. His main research interests include large model-powered visualization generation and evaluation, visual analytics, and explainable machine learning. For more information, please visit https://zekaishao25.github.io/.
\end{IEEEbiography}

\vspace{-21pt}

\begin{IEEEbiography}[{\includegraphics[width=1in,height=1.25in,clip,keepaspectratio]{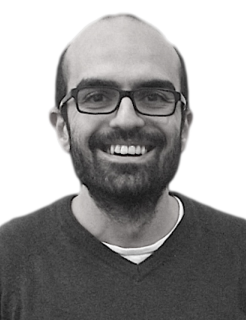}}]{Cagatay Turkay} is a Professor at the Centre for Interdisciplinary Methodologies at the University of Warwick, UK and a Turing Fellow at the Alan Turing Institute, London, UK. His research investigates the interactions between data, algorithms and people, and explores the role of interactive visualisation and other interaction mediums such as natural language  at this intersection. He frequently publishes his research on visualisation journals such as IEEE TVCG, CGF, and IEEE CG\&A, as well as journals in machine learning and data mining, and also recently co-authored a coursebook titled “Visual Analytics for Data Scientists’. He has been awarded the EuroVis Young Researcher 2019 award and named a EuroGraphics Junior Fellow in 2019.
\end{IEEEbiography}

\vspace{-21pt}

\begin{IEEEbiography}[{\includegraphics[width=1in,height=1.25in,clip,keepaspectratio]{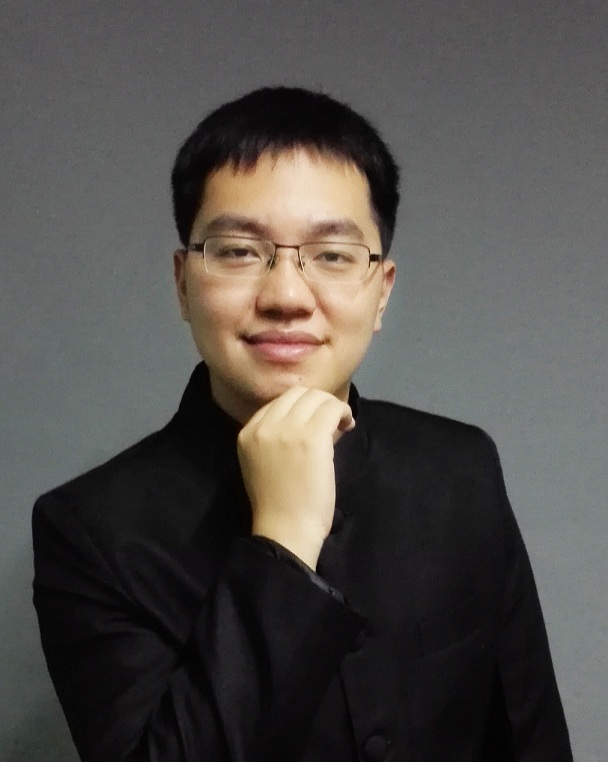}}]{Siming Chen} is an Associate Professor at School of Data Science, Fudan University. Prior to this, he was a Research Scientist at Fraunhofer Institute IAIS in Germany. He received his Ph.D. in computer science Peking University. His research interests are visualization and visual analytics, with the emphasis on Human-AI Collaboration, including LLM-driven visual analytics, social media and autonomous driving visual analytics. He has published 100 papers and more than 40 in top conferences and journals, including IEEE VIS, IEEE TVCG, EuroVis, ACM CHI, UIST, CSCW, etc. He served as multiple organizing chairs, committees and reviewers. He was awarded 10+ best paper/poster awards and honorable mentioned awards in multiple conferences For more information, please visit http://simingchen.me.
\end{IEEEbiography}

\fi

\end{document}